\documentclass[preprint]{aastex631}

\shorttitle{Verification of Hypervelocity Stars}
\shortauthors{Wojtkowska et al.}

\graphicspath{{./}{figures/}}

\begin{document}

\title{Verification of Hypervelocity Bulge Red Clump Stars}

\author{Gabriela Wojtkowska}
\affiliation{Astronomical Observatory, University of Warsaw, Al. Ujazdowskie 4 00-478 Warszawa, Poland}

\author{Radosław Poleski}
\affiliation{Astronomical Observatory, University of Warsaw, Al. Ujazdowskie 4 00-478 Warszawa, Poland}

\begin{abstract}

We verify candidate hypervelocity red clump stars located in the Galactic bulge that were selected based on the VVV and the $Gaia$ DR2 data by \citet{luna19}. To do so, we analyze data from the OGLE-IV survey: difference images and astrometric time series. We have data for 30 stars out of 34 hypervelocity candidates. We confirmed high proper motion of only one of these stars and find out that it is a nearby one, hence, not hypervelocity. To sum up, we do not confirm the candidate stars as hypervelocity ones. Hence, we disprove the  
production rate of hypervelocity red clump stars by the central supermassive black hole provided by \citet{luna19}.

\end{abstract}


\section{Introduction} \label{sec:intro}

Stars with galactocentric velocities higher than the Milky Way escape velocity are called hypervelocity stars \citep[HVSs; see][for a review]{brown15}. At the Sun position the escape velocity is estimated to be around $550~\mathrm{km\,s^{-1}}$, though the uncertainty is on the order of $10\%$ \citep{williams17,monari18,deason19}. The value changes with a galactocentric distance powered to approximately $-0.2$ \citep{williams17}. The existence of HVSs was first predicted by \citet{hills88} who also suggested, that HVSs might be used to prove that there is a supermassive black hole (BH) in the center of the Galaxy. The proof lies in the way HVSs might gain their velocities: if a binary encounters a BH, one of the stars enters the orbit around the BH, while the other one escapes and becomes an HVS. However, this is not the only possible mechanism in which an HVS might be formed. Two other possibilities of HVSs origin were suggested by \citet{yu03}. The first one is an interaction between two single stars in a dense region close to the Galactic center which results in one of the stars being ejected. The rate of ejected stars becoming an HVS is low ($10^{-11}$/yr). Another mechanism implies that there is an intermediate-mass BH orbiting around the supermassive BH in the Galactic Center. An interaction between a single star and this BH binary could also result in ejection of an HVS. There are also other theories regarding the HVSs origins, such as ejection from a supernova explosion \citep{abadi09}. 

The HVSs traveled dozens or hundreds of kpc, even if they are still young. 
Therefore, they can provide information about the orientation and shape of visible and dark matter in the Galaxy \citep{marchetti21}. Such studies can be more detailed if larger samples of HVSs are found. Currently there are around 40 well-established HVSs \citep{boubert18,elbadry23} though there were a number of studies verifying tens or even hundreds of candidates \citep[e.g.,][]{boubert18,li21,marchetti19,Ziegerer15}. 

In this paper, we verify HVS candidates presented by \citet{luna19}, who used the $Gaia$ Data Release 2 \citep[DR2;][]{Gaia18,2016A&A...595A...1G} and the VISTA Variables in the Via Lactea Survey \citep[VVV;][]{2010NewA...15..433M,vvv18} data to search for such stars toward the Galactic bulge. 
This search resulted in 34 bulge red clump (RC) stars that have high proper motions (PMs; here we consider $\mu > 100~\mathrm{mas\,yr^{-1}}$ as a high PM). The estimated heliocentric distances to these stars are between 6 and $11~\mathrm{kpc}$, hence, their transverse velocities are larger than the escape velocity at their location. Consequently, these 34 stars were considered candidate hypervelocity RC stars (HVRCSs) and if confirmed will almost double the sample of HVSs. Among these candidates, seven stars have PMs pointing away from the Galactic Center. Based on these seven stars, \citet{luna19} also estimated the total production rate of HVRCSs by the central supermassive
BH: $N\approx3.26 \times 10^{-4}~\mathrm{yr^{-1}}$.

We verify the \citet{luna19} candidates by measuring their PMs. Since the stars are relatively faint, for majority of them, the $Gaia$ astrometry has poor precision. Furthermore, because all of the stars are in the Bulge region, $Gaia$ is not able to collect data about every star every time it is observed, so it is hard to measure PMs. That is why we use the astrometry from the fourth phase of the Optical Gravitational Lensing Experiment \citep[OGLE-IV;][]{udalski15}. The OGLE-IV data used here contain hundreds of epochs collected over 8-9 years typically, hence, the PMs can be measured using these data reliably.

The paper is organized as follows: In Section \ref{sec:intro_luna} we introduce the methods used by \citet{luna19} and a list of the selected targets. We describe the data we used to verify the candidates in Section \ref{sec:data}. We present our methods in Section \ref{sec:methods} and results in Section \ref{sec:results}. These results are discussed in Section \ref{sec:summary}.

\section{Target selection} \label{sec:intro_luna}

\begin{table}[p]
\centering
\begin{tabular}{cccccc}
\hline
\hline
no. & $Gaia$ DR2 Source ID & $G$ [mag] & R.A. [$^{\circ}$] & Dec. [$^{\circ}$] & $d_\mathrm{L19}$ [kpc] \\
\hline
1 & 4043657695838288768 & $ 17.54 $ &  $ 268.26982081 $ & $ -31.75099406 $ & $ 8.89 $ \\
2 & 4050119216276193152 & $ 16.41 $ &  $ 272.29088035 $ & $ -29.66355549 $ & $ 8.34 $ \\
3 & 4050891554627761536 & $ 15.83 $ &  $ 273.35497285 $ & $ -28.05909984 $ & $ 7.44 $ \\
4 & 4053940195338701056 & $ 19.69 $ &  $ 265.66792433 $ & $ -32.97873039 $ & $ 9.06 $ \\
5 & 4055694466139424896 & $ 17.80 $ &  $ 268.30108179 $ & $ -31.31378523 $ & $ 7.12 $ \\
6 & 4055974493662728064 & $ 19.41 $ &  $ 267.27958008 $ & $ -30.60787786 $ & $ 7.39 $ \\
7 & 4056079565914165760 & $ 17.66 $ &  $ 268.45517554 $ & $ -30.94136746 $ & $ 10.28 $ \\
8 & 4056243191160985728 & $ 16.97 $ &  $ 269.92205870 $ & $ -29.96227506 $ & $ 9.93 $ \\
9 & 4056355405826018688 & $ 18.76 $ &  $ 267.68870103 $ & $ -30.39436359 $ & $ 7.79 $ \\
\textbf{ 10 } &  4056475218139133568 & $ 19.58 $ & $ 267.95180068 $ & $ -29.38624439 $ & $ 7.39 $ \\
11 & 4056562732513987200 & $ 17.69 $ &  $ 268.39043931 $ & $ -29.27377801 $ & $ 7.88 $ \\
12 & 4056575308188029056 & $ 18.19 $ &  $ 268.22054632 $ & $ -29.19157488 $ & $ 7.45 $ \\
13 & 4056799093014910848 & $ 18.30 $ &  $ 266.69167615 $ & $ -30.17967099 $ & $ 7.90 $ \\
14 & 4059583335633235456 & $ 18.07 $ &  $ 262.45358127 $ & $ -29.02761085 $ & $ 9.64 $ \\
\textbf{ 15 } &  4060422022558245248 & $ 18.66 $ & $ 264.26021020 $ & $ -28.50495081 $ & $ 7.41 $ \\
16 & 4060809836556331648 & $ 17.46 $ &  $ 266.36967922 $ & $ -27.13301289 $ & $ 10.08 $ \\
\textbf{ 17 } & 4060841348825794432 & $ 18.70 $ & $ 265.09103706 $ & $ -27.82013489 $ & $ 7.27 $ \\
18 & 4060858970977014784 & $ 18.91 $ &  $ 264.92158105 $ & $ -27.64125876 $ & $ 7.02 $ \\
\textbf{ 19 } & 4060875566737607936 & $ 19.00 $ & $ 265.43546801 $ & $ -27.52852279 $ & $ 7.32 $ \\
\textbf{ 20 } & 4061174432075039488 & $ 18.34 $ & $ 263.21342128 $ & $ -28.04755677 $ & $ 7.66 $ \\
21 & 4061331898466070528 & $ 18.39 $ &  $ 264.00263015 $ & $ -26.97930527 $ & $ 8.65 $ \\
\textbf{ 22 } & 4061839842728015360 & $ 18.58 $ & $ 265.12969259 $ & $ -26.23197613 $ & $ 10.71 $ \\
\textbf{ 23 } & 4062483636849326080 & $ 16.93 $ & $ 270.11054097 $ & $ -28.54237150 $ & $ 10.17 $ \\
24 & 4063159634713680384 & $ 16.78 $ &  $ 271.09154633 $ & $ -27.07320334 $ & $ 8.52 $ \\
25 & 4063179906906412032 & $ 16.98 $ &  $ 270.22075768 $ & $ -27.37100605 $ & $ 7.85 $ \\
26 & 4064174891897213312 & $ 19.74 $ &  $ 270.35669546 $ & $ -25.84324584 $ & $ 8.27 $ \\
27 & 4064524266009265024 & $ 17.24 $ &  $ 272.58165108 $ & $ -26.86186124 $ & $ 8.82 $ \\
28 & 4064649163703681792 & $ 17.56 $ &  $ 273.73570012 $ & $ -26.34858782 $ & $ 9.92 $ \\
29 & 4065639720555867904 & $ 18.11 $ &  $ 271.47611453 $ & $ -25.57422902 $ & $ 8.91 $ \\
30 & 4066241497024032768 & $ 18.12 $ &  $ 272.94310441 $ & $ -24.07029006 $ & $ 7.57 $ \\
31 & 4068160003069338112 & $ 18.00 $ &  $ 265.08148683 $ & $ -25.13218491 $ & $ 9.67 $ \\
32 & 4089677381280701312 & $ 16.81 $ &  $ 274.84724066 $ & $ -22.93497132 $ & $ 9.53 $ \\
33 & 4110276761609748096 & $ 17.75 $ &  $ 264.15273504 $ & $ -24.76831135 $ & $ 8.74 $ \\
34 & 4116316451998289664 & $ 18.45 $ &  $ 264.37354633 $ & $ -24.29284738 $ & $ 10.75 $ \\
\hline
\end{tabular}
\caption{List of 34 HVRCS candidates from \citet{luna19}. Precise equatorial coordinates are extracted based on the $Gaia$ DR2 identifiers. Seven final HVRCS candidates from \citet{luna19} are boldfaced in the first column. Distances (6th column) are taken from \citet{luna19}.}
\label{table_34}
\end{table}

\begin{table}[ht]
\centering
\begin{tabular}{ccccc}
\hline
\hline
no. & $Gaia$ DR2 Source ID & $d_\mathrm{L19}$ [kpc] & PM$_\mathrm{L19}$ [mas/yr] & $v_{T,\mathrm{L19}}$ [km/s] \\
\hline
10	&	4056475218139133568	&	7.39 &	$209	\pm 28$ &	7300 \\
15	&	4060422022558245248	&	7.41 &	$270	\pm 23$ &	9500 \\
17	&	4060841348825794432	&	7.27 &	$171	\pm 27$ &	5900 \\
19	&	4060875566737607936	&	7.32 &	$186	\pm 26$ &	6400 \\
20	&	4061174432075039488	&	7.66 &	$493	\pm 24$ &	17900 \\
22	&	4061839842728015360	&	10.71 &	$156	\pm 30$ &	7900 \\
23	&	4062483636849326080	&	10.17 &	$123	\pm 31$ &	5900 \\
\hline
\end{tabular}
\caption{Seven final HVRCS candidates selected by \citet{luna19}. Tangential velocities (last column) are based on PMs and distances from VVV.}
\label{table_luna}
\end{table}

\citet{luna19} analyzed the sample of 29,181,380 sources from the Bulge region using the VVV and $Gaia$ DR2 data. The sources had to appear in both datasets and have photometry in both near-IR and optical bands. Using the $G$-band magnitude from the $Gaia$ DR2 and the $K_s$-band magnitude from the VVV catalog, they created a color--magnitude diagram and selected stars belonging to the RC. Next, they measured PMs using two epochs: one from 2010 and one from 2015. To obtain the PM, the separation of the two centroids was measured and divided by the time difference. Further, they selected stars with high PMs (the limiting value was not indicated but all or allmost all selected stars have $\mu > 100~\mathrm{mas\,yr^{-1}}$ -- see their Tab.~3 and Fig.~2). After that, \citet{luna19} excluded nearby sources, defined as having a relative parallax uncertainty smaller than $20\%$ in the $Gaia$ DR2. Their research resulted in finding 34 stars with distances between 6 and 11 kpc and high PMs, listed in Table \ref{table_34}. 

If HVSs were created by the Hills mechanism, then they should be moving away from the Galactic Center. Therefore, \citet{luna19} also selected the sources with tangential velocities meeting this criterion. This selection resulted in a sample of seven stars, presented in Table \ref{table_luna}. 
Among the proper motions of these stars in \citet{luna19}, the smallest value is $123\pm31~\mathrm{mas\,yr^{-1}}$. This value should be compared with proper motion expected for a HVS in the Galactic bulge. For example, at a distance of 1~kpc from the Galactic center, the escape velocity whould be $840~\mathrm{km\,s^{-1}}$. If a star located 8~kpc from Sun has transverse of $840~\mathrm{km\,s^{-1}}$, then its proper motion is $21~\mathrm{mas\,yr^{-1}}$.

\section{Observational Data} \label{sec:data}

To verify the candidates proposed by \citet{luna19}, we used the $I$-band OGLE-IV astrometric data. The OGLE project started in 1992 with the original primary objective of searching for dark matter using the microlensing technique \citep{udalski93}. The project continued with four phases -- OGLE-I, OGLE-II, OGLE-III, and OGLE-IV \citep{udalski15}, differentiated by the camera and telescope upgrades. OGLE-IV phase started in 2010 and here we analyze data collected till March 2020. Observations are carried out with a 1.3-m telescope at the Las Campanas Observatory in Chile \citep{udalski97}. The telescope is equipped with a 32-chip 0.26"/pixel mosaic CCD camera. Its field of view is 1.4 $\mathrm{deg^2}$. The available astrometry has between 500 and 7000 epochs. We transform $(x, y)$ pixel coordinates measured on the individual images to the equatorial coordinates based on the bright stars observed by OGLE-IV and included in the $Gaia$ DR3 catalog \citep{2023A&A...674A...1G}. The PM and parallax values provided in the $Gaia$ DR3 were used to calculate equatorial coordinates for each stars at each epoch. These coordinates were then used to find the transformation.

Stars no. 6, 13, 24, and 26 were not analyzed, as they are not inside the OGLE-IV footprint or have very small number of epochs. We analyzed remaining 30 stars, which includes all seven stars from Table \ref{table_luna}.

We transformed the equatorial coordinates from Table~\ref{table_luna} to the pixel scale of the OGLE reference images, using software developed by the OGLE group\footnote{\url{http://ogle.astrouw.edu.pl/radec2field.html}}. We then searched the OGLE-IV catalog for nearest matching stars. The separation of equatorial coordinates between OGLE-IV and Gaia DR2 for almost all these stars is below $63~\mathrm{mas}$. The only exception is star no. 34 with a separation of $133~\mathrm{mas}$ (this star is later found to have highest proper motion in this sample). The coordinates are closely matched, hence, we the target stars are identified in the OGLE data securely. 

We provide $I$-band brightness of these stars in Table~\ref{propermotions}. In this sample, 22 stars are brighter than $I=17~\mathrm{mag}$ and all are brighter than $I=18.5~\mathrm{mag}$. Stars at these brightness limits have standard deviation of the OGLE time-series photometry of $\sigma_I < 0.03~\mathrm{mag}$ and $\sigma_I < 0.1~\mathrm{mag}$, respectively \citep{udalski15}.

\section{Methods} \label{sec:methods}

Below we first present qualitative analysis that is based on difference image analysis (DIA). Then we present quantitive analysis of centroids from point-spread function (PSF) method. The DIA and PSF analysis independently analyse OGLE-IV images.

\subsection{Qualitative analysis}

\begin{figure}[ht]
\centering
\includegraphics[width=0.49\textwidth]{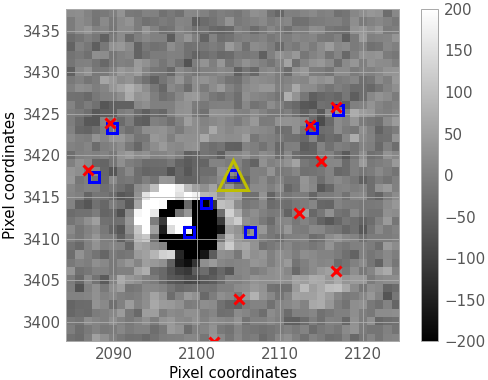} 
\includegraphics[width=0.49\textwidth]{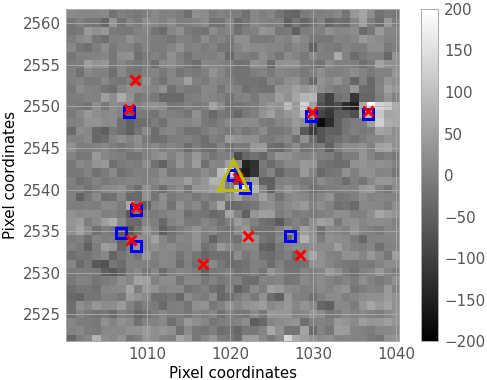} 
\includegraphics[width=0.49\textwidth]{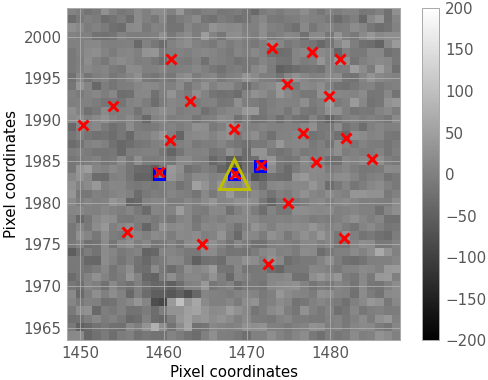} 
\includegraphics[width=0.49\textwidth]{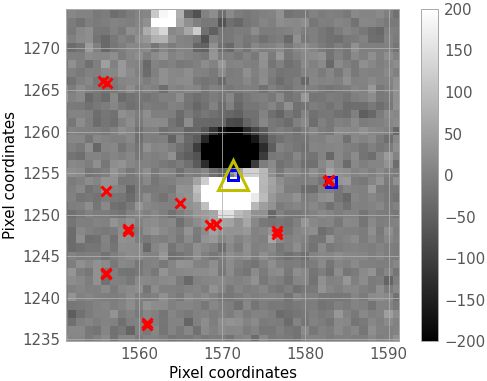} 

\vspace*{-13.5cm}
\hspace*{-3cm}
{\color{white} 
\textbf{Star no. 4} 
\hspace*{7cm}
\textbf{Star no. 8}
\vskip 6.5cm
\hspace*{-3cm}
\textbf{Star no. 10}
\hspace*{7cm}
\textbf{Star no. 34}
\vskip 6.5cm 
}

\caption{Difference images for stars no. 4, 8, 10, and 34 from the OGLE-IV survey. Both horizontal and vertical axis represent coordinates in pixel scale (1 pixel = 0.26 arcsec). The RA increases toward left and Dec. increases toward top.
The color bars plot flux in arbitrary units of the difference images. The target stars are marked with yellow triangles, the coordinates extracted from the $Gaia$ DR2 with solid blue squares, and the coordinates extracted from the VVV catalog with red crosses. Lower right panel shows the only high-proper motion star in among 30 verified targets.}
\label{diff}
\end{figure}

Our first step in verification of the \citet{luna19} HVRCS candidates was inspection of the OGLE-IV difference images. We used the provided $Gaia$ IDs and acquired the accurate positions of the targets as well as the stars within the 5 arcsec range from the $Gaia$ DR2\footnote{\url{https://vizier.u-strasbg.fr/viz-bin/VizieR-3?-source=I/345}} and VVV\footnote{\url{http://horus.roe.ac.uk/vsa/index.html}} catalogs. 
We present  difference images for four of HVRCS candidates in Figure \ref{diff}. These difference images compare the reference image constructed from the images from the begin of the OGLE-IV with an image taken under good seeing conditions at the end of analyzed time range. In difference images one can see how much the amount of flux changed since the beginning of OGLE-IV observations. If it stays the same, the pixel remains gray. If there were any significant changes, it turns white (more flux) or black (less flux). Therefore, if the star has a high PM, we should see it on the image as a white-black dipole \citep{eyer01}. 

Star no. 4 is an example of a star located near a very bright star, which is one of the factors that could affect the PM measurement. The image of star no. 8 highlights another possible problem. In the $Gaia$ DR2 catalog this star has a nearby companion, while in the VVV catalog the companion is not resolved. If the companion is resolved or not depends on the angular resolution of the telescope and, for ground-based telescopes, the seeing of given image. It is worth mentioning, the companion to the star no. 8 is detected in various surveys. The separation of these two stars  in the OGLE-III, OGLE-IV, and $Gaia$ DR3 catalogs are 0.55", 0.61", and 0.55", respectively. 
Most of the difference images looked similar to the one with star no. 10 -- there is no sign of the difference in flux coming from the area of the sky in which the star is located. Similarly to the star no. 8, the star no. 10 has a companion at a small separation (0.83", 0.84", and 0.88" based on OGLE-III, OGLE-IV, and $Gaia$ DR3, respectively). In our opinion, such nearby companions are the main reason of incorrect PM estimates. Inspection of difference images for all studied stars reveals that only one star has high PM: star no. 34.

\subsection{Quantitative analysis}

Next, we used the OGLE-IV astrometric time series to quantitatively verify HVRCS candidates. 
This time series 
come from the OGLE time-series astrometric database (OGLE-URANUS; Udalski et al. 2024, in prep.). In short, all OGLE images of the Galactic bulge fields with the seeing better than 1.25--1.35 arcsec (depending on stellar density of the field and the number of collected epochs) are reduced with the PSF-fitting software \texttt{DoPhot} \citep{schechter93} to derive precise stellar centroids in these dense fields. We note that \texttt{DoPhot} was already used to extract centroids and measure proper motions by other authors, e.g., \citet{2004MNRAS.348.1439S} and \citet{2012ApJ...745...42T}. 
Each image is divided into subfields covering half of a CCD chip and then these subfields are reduced separately. Centroids are extracted using slowly variable PSF. For each subfield, common stars between the OGLE-IV and a set of Gaia DR3 objects are found. After correction of the Gaia DR3 positions with the Gaia proper motions (to be on the same epoch as the current OGLE frame) the 5th order polynomial transformation of the OGLE current frame and (RA,DEC) of the Gaia reference frame is derived. Centroids of all detected on the current OGLE frame objects transformed to the Gaia reference frame form the OGLE astrometric time series.

We verified the reliability and accuracy of OGLE-URANUS by comparing proper motions to GAIA DR3. We selected stars from the same subfields as 30 targets verified here and limited the sample to the brightness range similar to that of target stars, i.e., $15.3~\mathrm{mag} < I < 17.4~\mathrm{mag}$. There were 156,000 stars in this sample and the dispersion of proper motion differences between OGLE-URANUS and GAIA DR3 were $1.7~\mathrm{mas\,yr^{-1}}$ in the R.A. component and $1.5~\mathrm{mas\,yr^{-1}}$ in the Dec. component. Hence, OGLE astrometric time-series are accurate enough to verify proper motions presented by \citet{luna19}.

\begin{figure}[p]
\centering
\includegraphics[width=\textwidth]{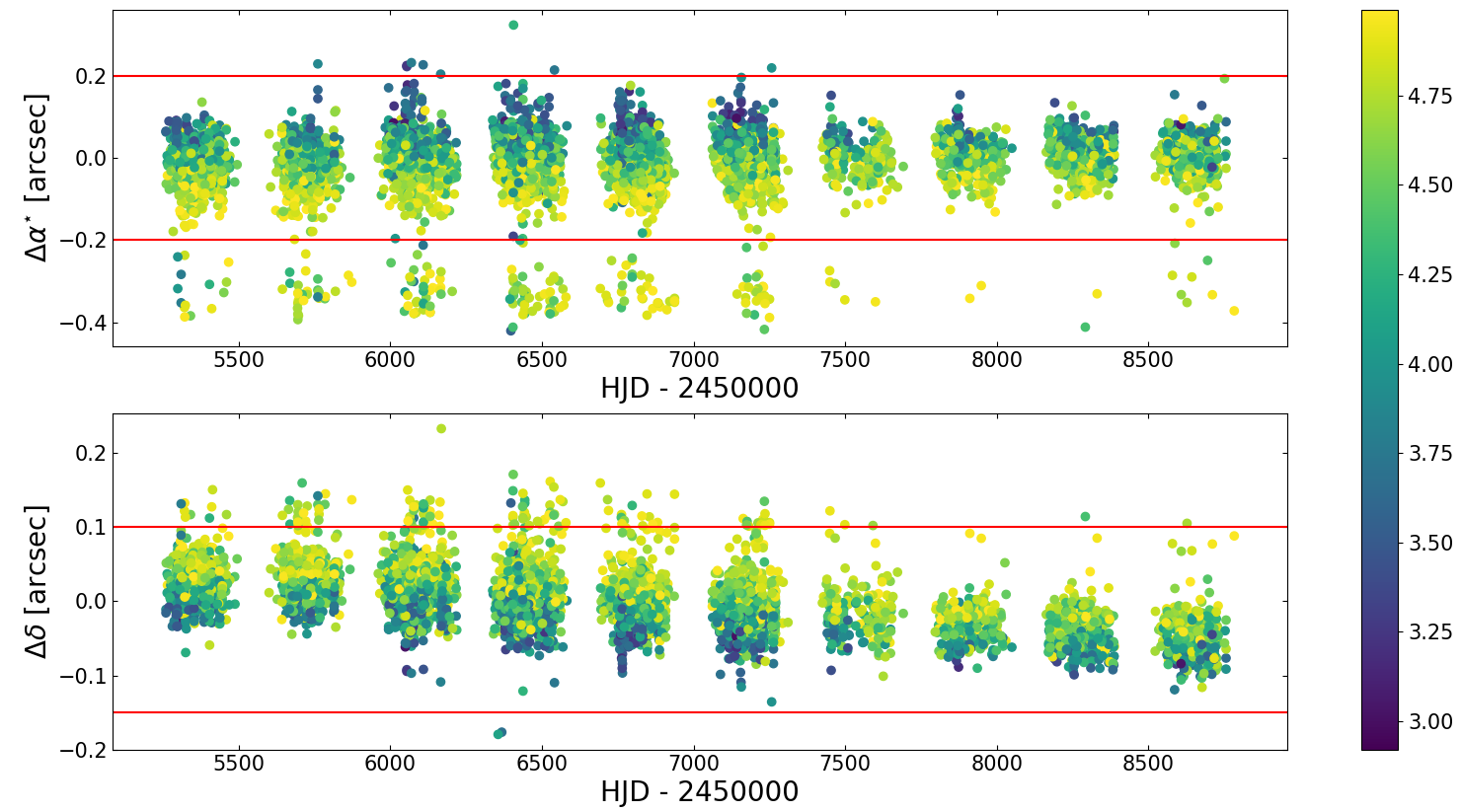}
\includegraphics[width=\textwidth]{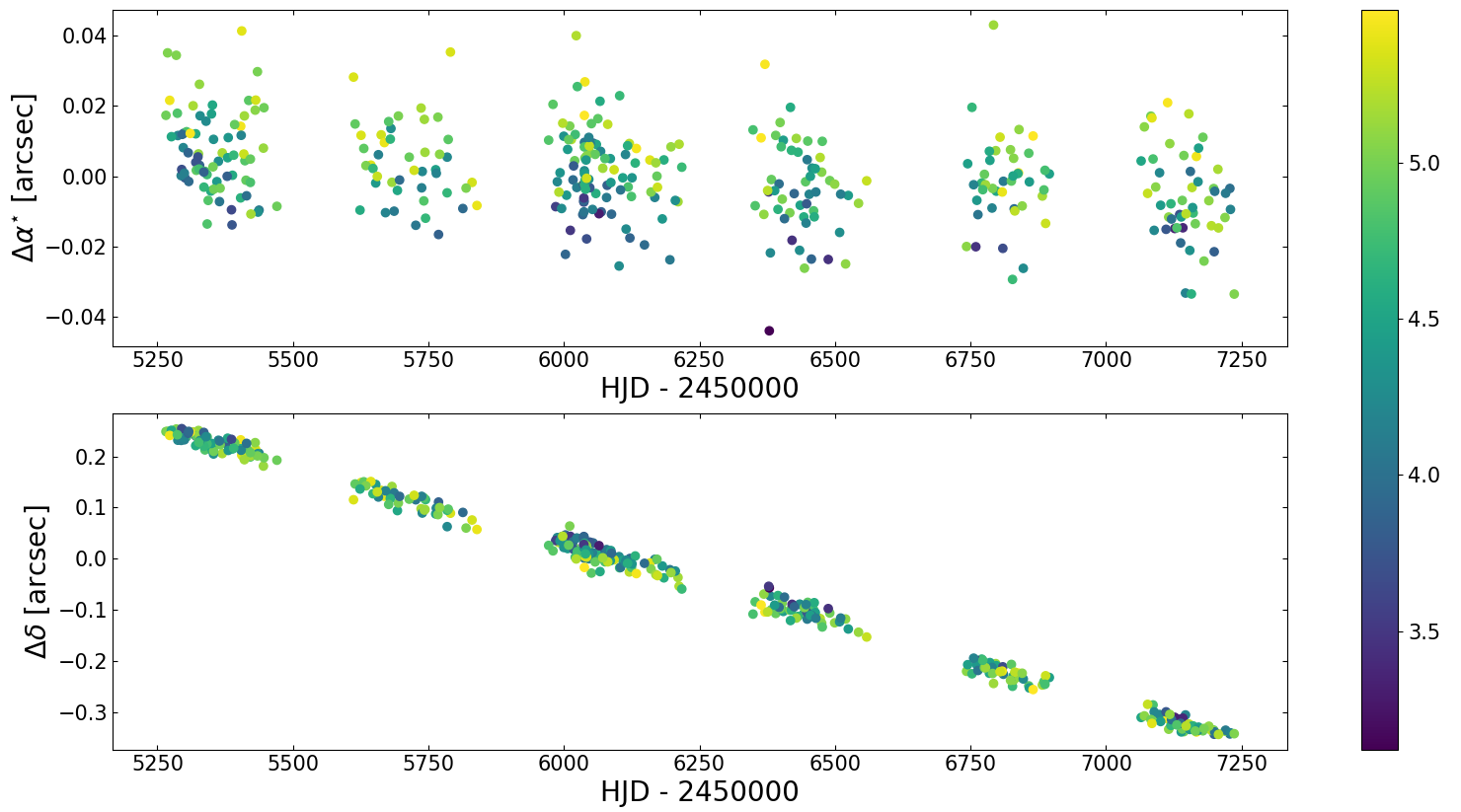}
\caption{Astrometry of stars no.~10 (top two panels) and 34 (bottom two panels) as a function of time. The Y axes are shifted for plotting purposes. The color-coding indicates seeing (in pixels). The data used for model fitting are between red horizontal lines. For the star no.~34, all data were used for fitting.
}
\label{xtyt}
\centering
\end{figure}

The astrometric time-series data consist of: right ascension ($\alpha$), declination ($\delta$), time ($t$), and seeing ($s$). We started the investigation by creating graphs visible in Figure~\ref{xtyt}. The horizontal axis represents time, and the vertical axes show the right ascension $\Delta\alpha^{\star}$ (upper graph) and the declination $\Delta\delta$ (lower graph). For all of the following images and equations, $\Delta\alpha^{\star}$ is corrected for projection effect:
\begin{equation}
    \Delta\alpha^{\star} = \Delta\alpha \cos{\delta}.
\end{equation}
The time-series astrometry of the analyzed stars shows outlying measurements which are caused by bad seeing, influence of nearby companion, asteroid etc. For each target star we selected epochs used for fitting based on visual inspection of plots analogous to Figure~\ref{xtyt}. The epochs used for fitting are between the red horizontal lines.

After that, we calculated the PMs of the stars by fitting a model to the differential astrometry:
\begin{equation}
        \Delta\alpha^{\star}(t) = \alpha_{0}+\mu_{\alpha}\,t+\pi \,p_{\alpha}(t),
\end{equation}
\begin{equation}
        \Delta\delta(t) = \delta_{0}+\mu_{\delta}\,t+\pi \,p_{\delta}(t),
        \label{42}
\end{equation}
where $\pi$ is the parallax, $\alpha_0$ and $\delta_0$ are the zero-points, $\mu_{\alpha}$ and $\mu_{\delta}$ are PMs, and $p_\alpha(t)$ and $p_\delta(t)$ project the parallax unit vector on $\alpha$ and $\delta$ directions, respectively. 

In order to sample $\mu_\alpha$, $\mu_\delta$, and $\pi$, we used the Markov Chain Monte Carlo (MCMC) method as implemented by \citet{Foreman13}. We define the likelihood function as:
\begin{equation}
        \ln{\mathcal{L}(\Delta\delta,\Delta\alpha^{\star}|\mu_{\delta},\mu_{\alpha},\delta_{0},\alpha_{0},\pi,f)}=-\frac{1}{2}\left(\chi^{2}_{\alpha}+\chi^{2}_{\delta}+\sum_{i=1}^{n}\ln{\sigma_i^2}\right), 
\end{equation}
where
\begin{equation}
        \chi^{2}_{\alpha}=\sum_{i=1}^{n}\left(\frac{\Delta\alpha^{\star}(t_{i})-\alpha_{0}-\mu_{\alpha}\,t_{i}-\pi \,p_{\alpha}(t_{i})}{\sigma_i}\right)^2,
\end{equation}
\begin{equation}
        \chi^{2}_{\delta}=\sum_{i=1}^{n}\left(\frac{\Delta\delta(t_{i})-\delta_{0}-\mu_{\delta}\,t_{i}-\pi \,p_{\delta}(t_{i})}{\sigma_i}\right)^2.
\end{equation}
The uncertainties were obtained by multiplying seeing by a constant \citep{kuijken02}:
\begin{equation}
    \sigma_i = f s_i,
\end{equation}
where $f$ is the fitted multiplication constant.

\section{Results} \label{sec:results}

\begin{figure}[ht]
\centering
\includegraphics[width=.95\textwidth]{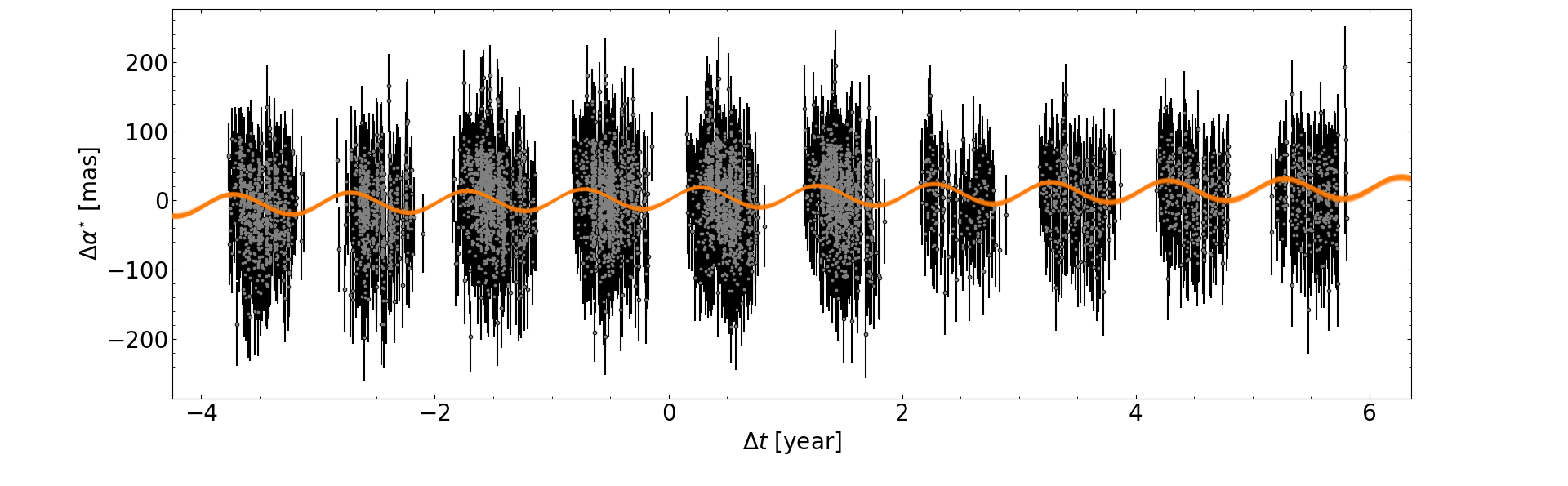}
\includegraphics[width=.95\textwidth]{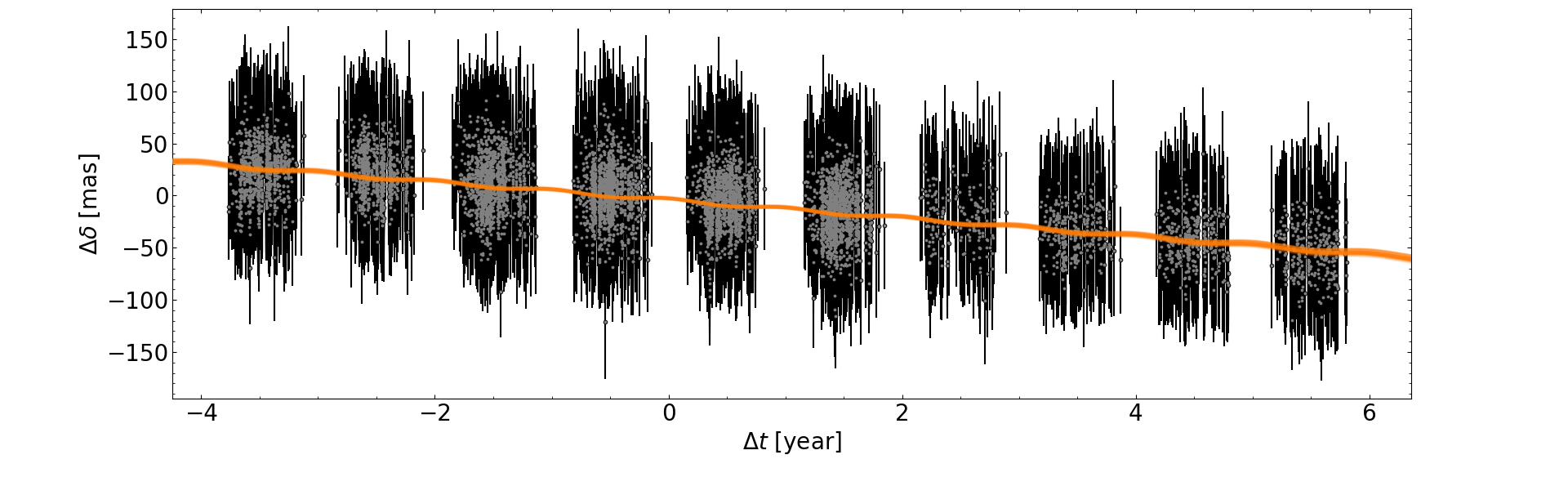}
\includegraphics[width=.95\textwidth]{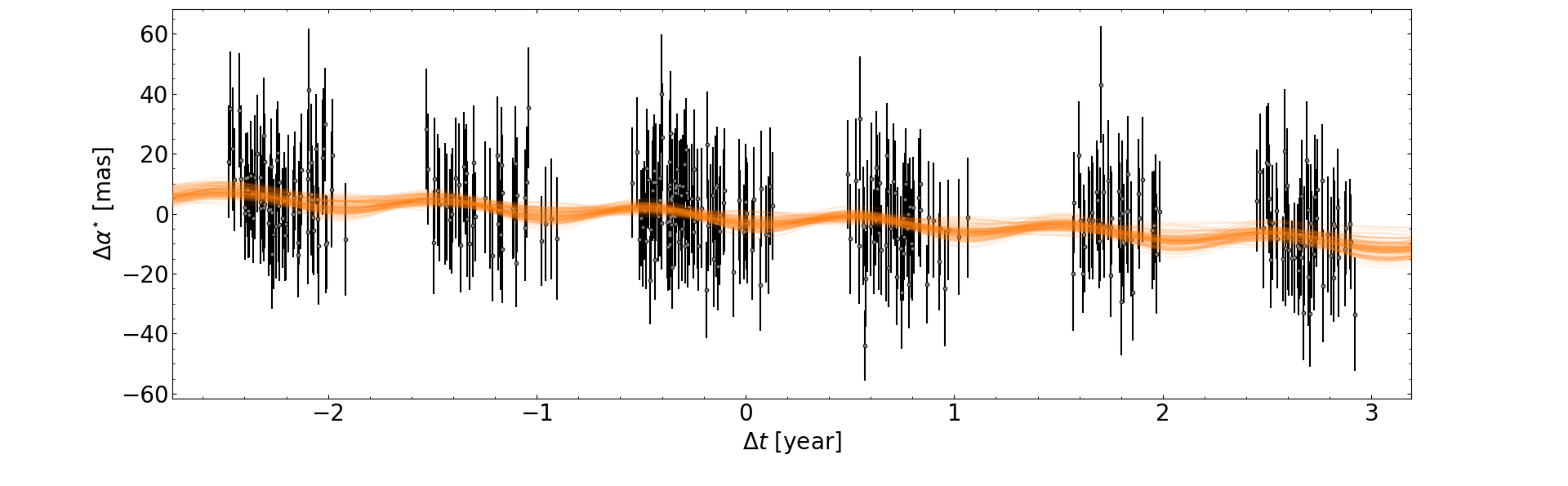}
\includegraphics[width=.95\textwidth]{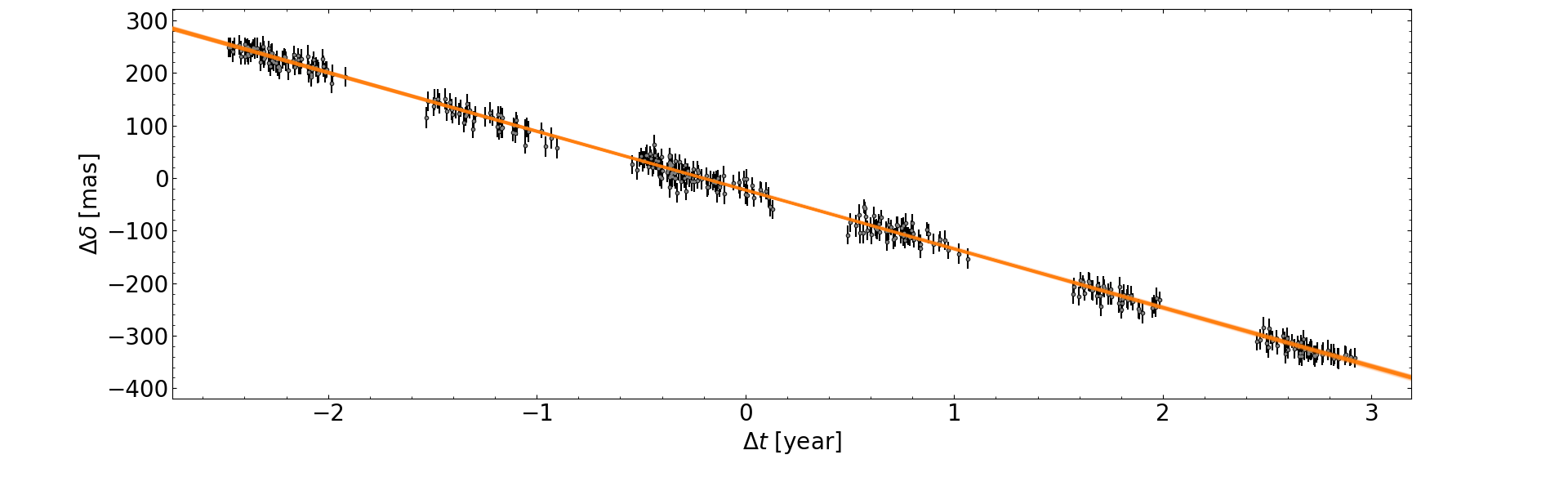}
\caption{Differential astrometry vs. time with fitted models (orange curves) for stars no. 10 (top) and 34 (bottom).}
\label{pmdelta}
\centering
\end{figure}

\begin{figure}[p]
\centering
\includegraphics[width=19cm]{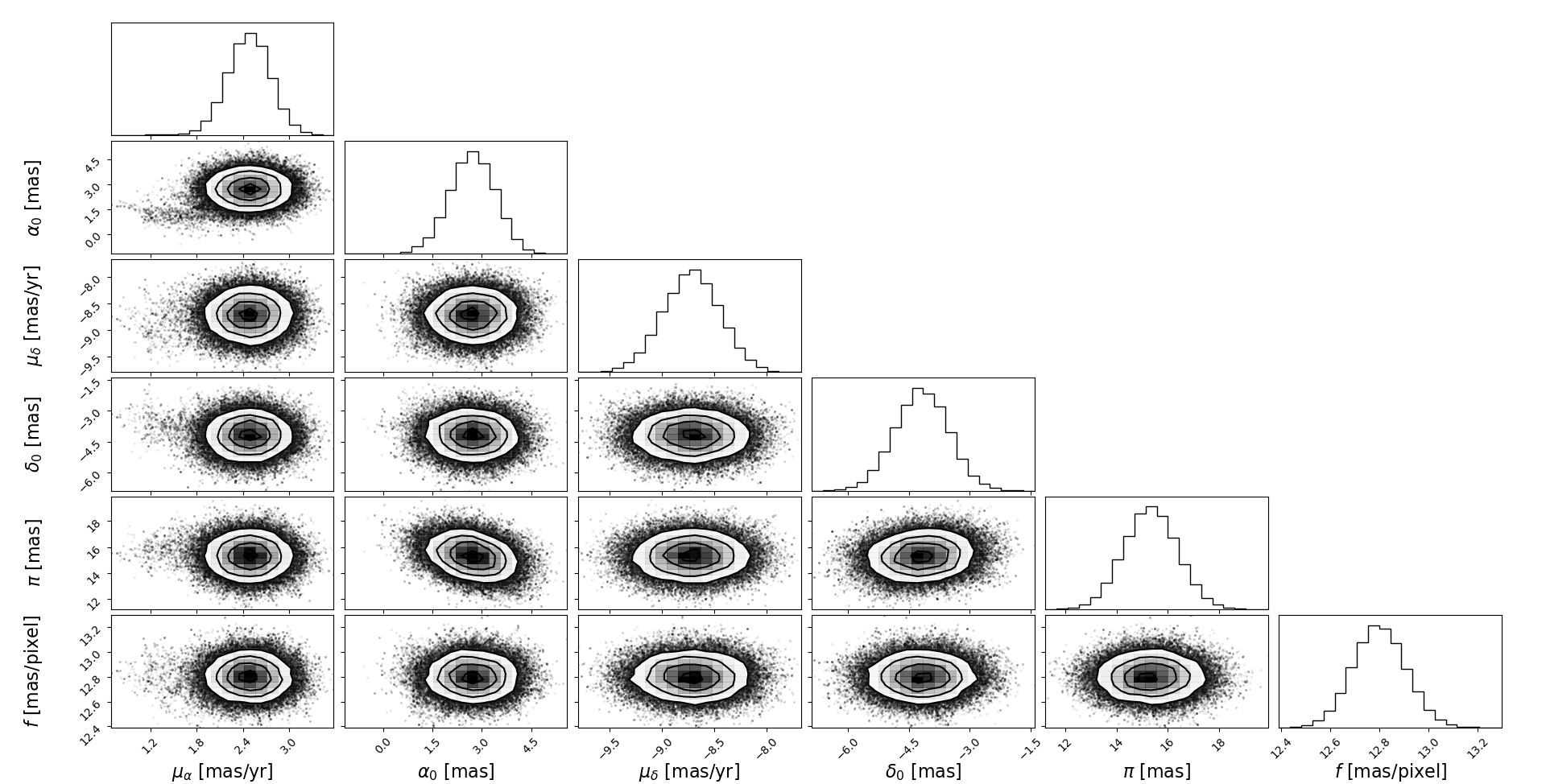}
\includegraphics[width=19cm]{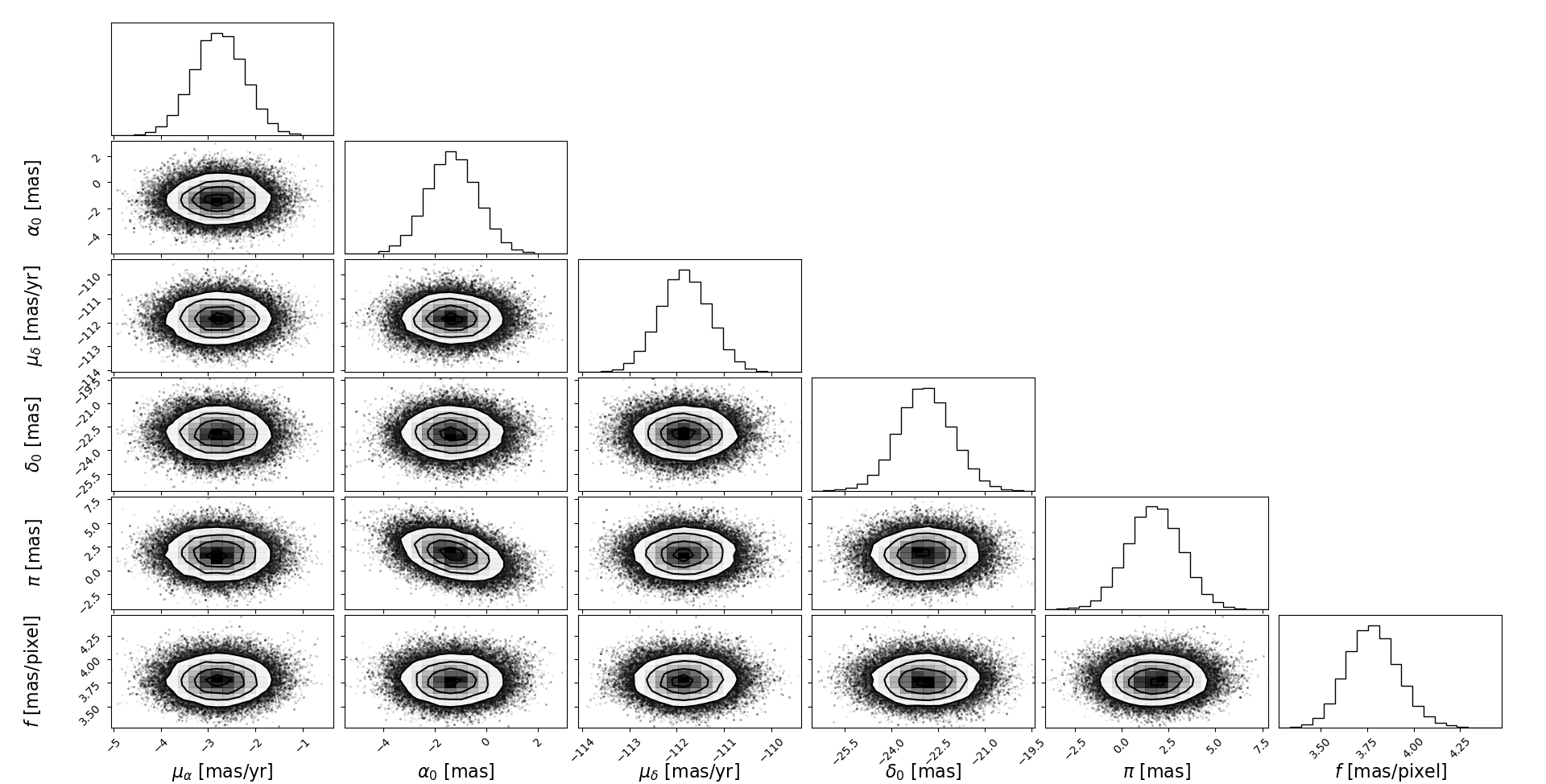}
\caption{Corner plots of posterior distribution of parameters for stars no. 10 (top) and 34 (bottom).}
\label{corner}
\centering
\end{figure}

We present the calculated PMs for all the 30 stars in Table \ref{propermotions}. Based on these results, only star no. 34 has a high PM. For illustration, we generated the plots showing differential astrometry and fitted models -- see Figure \ref{pmdelta} for examples. Again, we can see a clear difference between the star with low (star no. 10) and high (star no. 34) PM. We also present the corner plots for the same stars in Figure \ref{corner}. It is visible, that all of the parameters are independent of each other. 

Given all the information gathered, only star no. 34 can be considered a high-PM star. This star was not chosen by \citet{luna19} as one of the seven HVRCSs moving away from the Galactic Center because of its PM direction. However, this star is not an HVS but a nearby star instead. \citet{luna19} did not classify it as a nearby star because its $Gaia$ DR2 parallax of $1.03\pm0.64~\mathrm{mas}$ is consistent with 0. In the $Gaia$ DR3 \citep[published after][]{luna19}, the parallax is larger and its uncertainty is smaller: $1.59\pm0.31~\mathrm{mas}$ (the goodness-of-fit parameter RUWE is 1.477, which is slightly larger than the frequently assumed limit of 1.4). This parallax combined with the photogeometric prior places the star at a distance of $686^{+71}_{-100}~\mathrm{pc}$ \citep{Bailer-Jones21}. 
One can calculate the transverse velocity by multiplying PM and distance. 
Our measurement of PM and $Gaia$ distance result in transverse velocity of $364^{+38}_{-53}~\mathrm{km/s}$ which reveals a star bound to the Galaxy.

Some of the parallax uncertainties in Table~\ref{propermotions} are small, hence, we tried to estimate if our data analysis procedures could underestimate the parallax uncertainty. We assumed that the parallax is underestimated by a constant that is added in squares, i.e., $\sigma'_{\pi} = \sqrt{\sigma_\pi^2+\epsilon^2}$. We compare the parallaxes and their uncertainties derived from the OGLE-IV astrometry to the distances derived by \citet{Bailer-Jones21} from the $Gaia$ DR3 astrometry and informative priors. The comparison was based on methods presented by \citet{Luri2018} and eight analysed stars which have well-measured $Gaia$ data (i.e., goodness-of-fit parameter RUWE smaller than 1.4; these are stars no. 2, 4, 9, 11, 14, 16, 28, and 30). We obtained $\epsilon = 1.77^{+0.77}_{-0.48}~\mathrm{mas}$. We note that this value was derived based on a small sample of red clump stars and some of them have nearby companions.

\begin{table}
\centering
\begin{tabular}{cccccc}
\hline\hline
no.    & $I$ [mag] &  $\mu_{\alpha}~\mathrm{[mas/yr]}$  & $\mu_{\delta}~\mathrm{[mas/yr]}$  & $\pi~\mathrm{[mas]}$  &    $f$ [mas/pixel] \\ \hline
1 & $16.23$ & $-5.61\pm0.55$ & $-2.03\pm0.54$ & $-4.6\pm2.3$ & $10.92\pm0.25$ \\
2 & $15.36$ & $-2.10\pm0.27$ & $-5.32\pm0.26$ & $-3.1\pm1.1$ & $5.07\pm0.12$ \\
3 & $14.78$ & $1.59\pm0.37$ & $-1.79\pm0.37$ & $-2.2\pm1.1$ & $3.62\pm0.12$ \\
4 & $17.86$ & $0.8\pm1.2$ & $-2.8\pm1.3$ & $6.1\pm5.5$ & $16.41\pm0.55$ \\
5 & $16.47$ & $-3.087\pm0.075$ & $0.810\pm0.074$ & $0.78\pm0.31$ & $2.815\pm0.034$ \\
7 & $16.38$ & $-9.21\pm0.16$ & $-4.25\pm0.16$ & $5.42\pm0.66$ & $5.779\pm0.073$ \\
8 & $15.51$ & $0.88\pm0.45$ & $-3.65\pm0.45$ & $-2.0\pm1.9$ & $11.75\pm0.20$ \\
9 & $17.26$ & $-1.42\pm0.17$ & $-5.38\pm0.18$ & $2.92\pm0.69$ & $8.503\pm0.072$ \\
\textbf{10} & $17.99$ & $2.48\pm0.27$ & $-8.73\pm0.27$ & $15.3\pm1.0$ & $12.80\pm0.11$ \\
11 & $16.30$ & $1.23\pm0.31$ & $-1.33\pm0.32$ & $-0.6\pm1.2$ & $9.99\pm0.14$ \\
12 & $16.69$ & $0.77\pm0.24$ & $-4.04\pm0.24$ & $-5.1\pm1.1$ & $8.21\pm0.11$ \\
14 & $16.74$ & $-3.32\pm0.44$ & $-8.24\pm0.44$ & $-5.9\pm1.9$ & $7.61\pm0.21$ \\
\textbf{15} & $17.17$ & $-1.49\pm0.50$ & $-0.77\pm0.50$ & $-8.2\pm1.8$ & $7.72\pm0.20$ \\
16 & $16.15$ & $-0.23\pm0.15$ & $0.22\pm0.14$ & $-0.85\pm0.65$ & $2.244\pm0.066$ \\
\textbf{17} & $17.22$ & $-0.34\pm0.62$ & $-9.07\pm0.61$ & $-1.8\pm2.0$ & $9.85\pm0.22$ \\
18 & $17.37$ & $-1.1\pm1.1$ & $-2.4\pm1.0$ & $-1.5\pm3.5$ & $12.38\pm0.41$ \\
\textbf{19} & $17.53$ & $-1.59\pm0.69$ & $-8.87\pm0.69$ & $-7.1\pm2.2$ & $11.42\pm0.24$ \\
\textbf{20} & $16.84$ & $-4.94\pm0.66$ & $-5.46\pm0.63$ & $-5.7\pm2.6$ & $8.84\pm0.28$ \\
21 & $16.97$ & $-3.11\pm0.20$ & $-7.35\pm0.20$ & $9.3\pm0.8$ & $7.61\pm0.09$ \\
\textbf{22} & $18.43$ & $-8.42\pm0.27$ & $-0.96\pm0.27$ & $-1.7\pm1.1$ & $4.03\pm0.12$ \\
\textbf{23} & $15.75$ & $-2.34\pm0.51$ & $-7.13\pm0.51$ & $-4.1\pm1.9$ & $10.72\pm0.20$ \\
25 & $15.67$ & $-2.80\pm0.18$ & $-5.16\pm0.17$ & $-1.16\pm0.75$ & $5.886\pm0.078$ \\
27 & $15.93$ & $1.16\pm0.60$ & $-6.04\pm0.60$ & $-6.5\pm2.7$ & $8.40\pm0.27$ \\
28 & $16.16$ & $-2.2\pm1.0$ & $-5.1\pm1.0$ & $1.1\pm4.3$ & $8.04\pm0.50$ \\
29 & $16.59$ & $-2.70\pm0.87$ & $-2.61\pm0.88$ & $0.4\pm3.5$ & $15.26\pm0.38$ \\
30 & $16.74$ & $-2.43\pm0.86$ & $-9.36\pm0.85$ & $-0.7\pm2.7$ & $6.00\pm0.26$ \\
31 & $16.65$ & $-0.2\pm1.4$ & $-5.4\pm1.4$ & $-3.7\pm3.2$ & $6.86\pm0.36$ \\
32 & $15.42$ & $-6.7\pm1.6$ & $-7.2\pm1.6$ & $18.4\pm5.3$ & $13.42\pm0.52$ \\
33 & $16.40$ & $-1.4\pm4.6$ & $-3.9\pm4.6$ & $26.0\pm17.0$ & $47.0\pm1.8$ \\
34 & $16.91$ & $-2.80\pm0.57$ & $-111.84\pm0.54$ & $1.8\pm1.3$ & $3.78\pm0.14$ \\
\hline
\end{tabular}
\caption{Proper motions, parallaxes, and uncertainty to seeing ratios for the 30 analyzed stars. The second column gives the $I$-band brightness in the OGLE-IV photometry. The other values are derived from the OGLE-IV time-series astrometry. Seven final HVRCS candidates from \citet{luna19} are boldfaced in the first column.
}
\label{propermotions}
\end{table}

\section{Discussion} \label{sec:summary}

Our verification does not confirm, that the candidates provided by \citet{luna19} are HVSs. Only one of these stars has a high PM but it is a nearby object, not a bulge one, hence, not a hypervelocity one. We analyzed 30 out of 34 HVRCSs including all seven stars which \citet{luna19} used to estimate the production rate of HVRCSs by the central supermassive black hole. Thus, this rate is not supported by any evidence.

In our opinion, the main reason why \citet{luna19} results are so different from ours is the approach of \citet{luna19} to measure the PMs based on only two epochs. In a very large sample of stars, even a very rare case of measuring one of these positions incorrectly may produce a small sample (about 30 objects in the case discussed here) of stars with high and incorrect PMs. These incorrect measurements of positions are mostly caused by neighboring stars that affect the centroid of the investigated stars. 

We note that the tangential velocities calculated by \citet[][their Table 3 presented here as Table 2]{luna19} are extremely large. All of them are at least three times larger than the highest 3D velocity of a well-established HVS \citep[$1,755~\mathrm{km/s}$;][]{koposov20}. 

\begin{acknowledgments}
We are grateful to the OGLE team (led by A. Udalski) for providing their data. 
This paper is based on the M.Sc. thesis written by G.W. We thank anonymous reviewer for suggestions that improved the text.
We also thank Alonso Luna for comments on a draft.
This work has made use of data from the European Space Agency (ESA) mission {\it Gaia} (\url{https://www.cosmos.esa.int/gaia}), processed by the {\it Gaia} Data Processing and Analysis Consortium (DPAC, \url{https://www.cosmos.esa.int/web/gaia/dpac/consortium}). Funding for the DPAC
has been provided by national institutions, in particular the institutions participating in the {\it Gaia} Multilateral Agreement.
\end{acknowledgments}

\vspace{5mm}
\facilities{1.3-m Warsaw Telescope at the Las Campanas Observatory, \textit{Gaia} satellite }

\software{
\texttt{Astropy} \citep{astropy:2013, astropy:2018, astropy:2022},
\texttt{corner.py} \citep{corner},
\texttt{DoPhot} \citep{schechter93}, 
\texttt{EMCEE} \citep{emcee},
\texttt{matplotlib} \citep{Hunter07}, 
\texttt{numpy} \citep{harris2020array}, 
OGLE databases \citep{Szymanski93}.
} 

\bibliography{publikacja}{}

\begin{thebibliography}{}
\expandafter\ifx\csname natexlab\endcsname\relax\def\natexlab#1{#1}\fi
\providecommand{\url}[1]{\href{#1}{#1}}
\providecommand{\dodoi}[1]{doi:~\href{http://doi.org/#1}{\nolinkurl{#1}}}
\providecommand{\doeprint}[1]{\href{http://ascl.net/#1}{\nolinkurl{http://ascl.net/#1}}}
\providecommand{\doarXiv}[1]{\href{https://arxiv.org/abs/#1}{\nolinkurl{https://arxiv.org/abs/#1}}}

\bibitem[{{Abadi} {et~al.}(2009){Abadi}, {Navarro}, \& {Steinmetz}}]{abadi09}
{Abadi}, M.~G., {Navarro}, J.~F., \& {Steinmetz}, M. 2009, \apjl, 691, L63.
\newblock \doarXiv{0810.1429}

\bibitem[{{Alonso-Garc{\'\i}a} {et~al.}(2018){Alonso-Garc{\'\i}a}, {Saito}, {Hempel}, {Minniti}, {Pullen}, {Catelan}, {Ramos}, {Cross}, {Gonzalez}, {Lucas}, {Palma}, {Valenti}, \& {Zoccali}}]{vvv18}
{Alonso-Garc{\'\i}a}, J., {Saito}, R.~K., {Hempel}, M., {et~al.} 2018, \aap, 619, A4.
\newblock \doarXiv{1808.06139}

\bibitem[{{Astropy Collaboration} {et~al.}(2013){Astropy Collaboration}, {Robitaille}, {Tollerud}, {Greenfield}, {Droettboom}, {Bray}, {Aldcroft}, {Davis}, {Ginsburg}, {Price-Whelan}, {Kerzendorf}, {Conley}, {Crighton}, {Barbary}, {Muna}, {Ferguson}, {Grollier}, {Parikh}, {Nair}, {Unther}, {Deil}, {Woillez}, {Conseil}, {Kramer}, {Turner}, {Singer}, {Fox}, {Weaver}, {Zabalza}, {Edwards}, {Azalee Bostroem}, {Burke}, {Casey}, {Crawford}, {Dencheva}, {Ely}, {Jenness}, {Labrie}, {Lim}, {Pierfederici}, {Pontzen}, {Ptak}, {Refsdal}, {Servillat}, \& {Streicher}}]{astropy:2013}
{Astropy Collaboration}, {Robitaille}, T.~P., {Tollerud}, E.~J., {et~al.} 2013, \aap, 558, A33, \dodoi{10.1051/0004-6361/201322068}

\bibitem[{{Astropy Collaboration} {et~al.}(2018){Astropy Collaboration}, {Price-Whelan}, {Sip{\H{o}}cz}, {G{\"u}nther}, {Lim}, {Crawford}, {Conseil}, {Shupe}, {Craig}, {Dencheva}, {Ginsburg}, {Vand erPlas}, {Bradley}, {P{\'e}rez-Su{\'a}rez}, {de Val-Borro}, {Aldcroft}, {Cruz}, {Robitaille}, {Tollerud}, {Ardelean}, {Babej}, {Bach}, {Bachetti}, {Bakanov}, {Bamford}, {Barentsen}, {Barmby}, {Baumbach}, {Berry}, {Biscani}, {Boquien}, {Bostroem}, {Bouma}, {Brammer}, {Bray}, {Breytenbach}, {Buddelmeijer}, {Burke}, {Calderone}, {Cano Rodr{\'\i}guez}, {Cara}, {Cardoso}, {Cheedella}, {Copin}, {Corrales}, {Crichton}, {D'Avella}, {Deil}, {Depagne}, {Dietrich}, {Donath}, {Droettboom}, {Earl}, {Erben}, {Fabbro}, {Ferreira}, {Finethy}, {Fox}, {Garrison}, {Gibbons}, {Goldstein}, {Gommers}, {Greco}, {Greenfield}, {Groener}, {Grollier}, {Hagen}, {Hirst}, {Homeier}, {Horton}, {Hosseinzadeh}, {Hu}, {Hunkeler}, {Ivezi{\'c}}, {Jain}, {Jenness}, {Kanarek}, {Kendrew}, {Kern}, {Kerzendorf}, {Khvalko}, {King}, {Kirkby}, {Kulkarni},
  {Kumar}, {Lee}, {Lenz}, {Littlefair}, {Ma}, {Macleod}, {Mastropietro}, {McCully}, {Montagnac}, {Morris}, {Mueller}, {Mumford}, {Muna}, {Murphy}, {Nelson}, {Nguyen}, {Ninan}, {N{\"o}the}, {Ogaz}, {Oh}, {Parejko}, {Parley}, {Pascual}, {Patil}, {Patil}, {Plunkett}, {Prochaska}, {Rastogi}, {Reddy Janga}, {Sabater}, {Sakurikar}, {Seifert}, {Sherbert}, {Sherwood-Taylor}, {Shih}, {Sick}, {Silbiger}, {Singanamalla}, {Singer}, {Sladen}, {Sooley}, {Sornarajah}, {Streicher}, {Teuben}, {Thomas}, {Tremblay}, {Turner}, {Terr{\'o}n}, {van Kerkwijk}, {de la Vega}, {Watkins}, {Weaver}, {Whitmore}, {Woillez}, {Zabalza}, \& {Astropy Contributors}}]{astropy:2018}
{Astropy Collaboration}, {Price-Whelan}, A.~M., {Sip{\H{o}}cz}, B.~M., {et~al.} 2018, \aj, 156, 123, \dodoi{10.3847/1538-3881/aabc4f}

\bibitem[{{Astropy Collaboration} {et~al.}(2022){Astropy Collaboration}, {Price-Whelan}, {Lim}, {Earl}, {Starkman}, {Bradley}, {Shupe}, {Patil}, {Corrales}, {Brasseur}, {N{"o}the}, {Donath}, {Tollerud}, {Morris}, {Ginsburg}, {Vaher}, {Weaver}, {Tocknell}, {Jamieson}, {van Kerkwijk}, {Robitaille}, {Merry}, {Bachetti}, {G{"u}nther}, {Aldcroft}, {Alvarado-Montes}, {Archibald}, {B{'o}di}, {Bapat}, {Barentsen}, {Baz{'a}n}, {Biswas}, {Boquien}, {Burke}, {Cara}, {Cara}, {Conroy}, {Conseil}, {Craig}, {Cross}, {Cruz}, {D'Eugenio}, {Dencheva}, {Devillepoix}, {Dietrich}, {Eigenbrot}, {Erben}, {Ferreira}, {Foreman-Mackey}, {Fox}, {Freij}, {Garg}, {Geda}, {Glattly}, {Gondhalekar}, {Gordon}, {Grant}, {Greenfield}, {Groener}, {Guest}, {Gurovich}, {Handberg}, {Hart}, {Hatfield-Dodds}, {Homeier}, {Hosseinzadeh}, {Jenness}, {Jones}, {Joseph}, {Kalmbach}, {Karamehmetoglu}, {Ka{l}uszy{'n}ski}, {Kelley}, {Kern}, {Kerzendorf}, {Koch}, {Kulumani}, {Lee}, {Ly}, {Ma}, {MacBride}, {Maljaars}, {Muna}, {Murphy}, {Norman}, {O'Steen},
  {Oman}, {Pacifici}, {Pascual}, {Pascual-Granado}, {Patil}, {Perren}, {Pickering}, {Rastogi}, {Roulston}, {Ryan}, {Rykoff}, {Sabater}, {Sakurikar}, {Salgado}, {Sanghi}, {Saunders}, {Savchenko}, {Schwardt}, {Seifert-Eckert}, {Shih}, {Jain}, {Shukla}, {Sick}, {Simpson}, {Singanamalla}, {Singer}, {Singhal}, {Sinha}, {Sip{H{o}}cz}, {Spitler}, {Stansby}, {Streicher}, {{{S}}umak}, {Swinbank}, {Taranu}, {Tewary}, {Tremblay}, {Val-Borro}, {Van Kooten}, {Vasovi{'c}}, {Verma}, {de Miranda Cardoso}, {Williams}, {Wilson}, {Winkel}, {Wood-Vasey}, {Xue}, {Yoachim}, {Zhang}, {Zonca}, \& {Astropy Project Contributors}}]{astropy:2022}
{Astropy Collaboration}, {Price-Whelan}, A.~M., {Lim}, P.~L., {et~al.} 2022, \apj, 935, 167, \dodoi{10.3847/1538-4357/ac7c74}

\bibitem[{{Bailer-Jones} {et~al.}(2021){Bailer-Jones}, {Rybizki}, {Fouesneau}, {Demleitner}, \& {Andrae}}]{Bailer-Jones21}
{Bailer-Jones}, C.~A.~L., {Rybizki}, J., {Fouesneau}, M., {Demleitner}, M., \& {Andrae}, R. 2021, \aj, 161, 147, \dodoi{10.3847/1538-3881/abd806}

\bibitem[{{Boubert} {et~al.}(2018){Boubert}, {Guillochon}, {Hawkins}, {Ginsburg}, {Evans}, \& {Strader}}]{boubert18}
{Boubert}, D., {Guillochon}, J., {Hawkins}, K., {et~al.} 2018, \mnras, 479, 2789.
\newblock \doarXiv{1804.10179}

\bibitem[{{Brown}(2015)}]{brown15}
{Brown}, W.~R. 2015, \araa, 53, 15, \dodoi{10.1146/annurev-astro-082214-122230}

\bibitem[{{Deason} {et~al.}(2019){Deason}, {Fattahi}, {Belokurov}, {Evans}, {Grand}, {Marinacci}, \& {Pakmor}}]{deason19}
{Deason}, A.~J., {Fattahi}, A., {Belokurov}, V., {et~al.} 2019, \mnras, 485, 3514, \dodoi{10.1093/mnras/stz623}

\bibitem[{{El-Badry} {et~al.}(2023){El-Badry}, {Shen}, {Chandra}, {Bauer}, {Fuller}, {Strader}, {Chomiuk}, {Naidu}, {Caiazzo}, {Rodriguez}, {Nagarajan}, {Yamaguchi}, {Vanderbosch}, {Roulston}, {G{\"a}nsicke}, {Han}, {Burdge}, {Filippenko}, {Brink}, \& {Zheng}}]{elbadry23}
{El-Badry}, K., {Shen}, K.~J., {Chandra}, V., {et~al.} 2023, The Open Journal of Astrophysics, 6, 28, \dodoi{10.21105/astro.2306.03914}

\bibitem[{{Eyer} \& {Wo{\'z}niak}(2001)}]{eyer01}
{Eyer}, L., \& {Wo{\'z}niak}, P.~R. 2001, \mnras, 327, 601.
\newblock \doarXiv{astro-ph/0102027}

\bibitem[{Foreman-Mackey(2016)}]{corner}
Foreman-Mackey, D. 2016, The Journal of Open Source Software, 1, 24, \dodoi{10.21105/joss.00024}

\bibitem[{{Foreman-Mackey} {et~al.}(2013{\natexlab{a}}){Foreman-Mackey}, {Hogg}, {Lang}, \& {Goodman}}]{Foreman13}
{Foreman-Mackey}, D., {Hogg}, D.~W., {Lang}, D., \& {Goodman}, J. 2013{\natexlab{a}}, \pasp, 125, 306.
\newblock \doarXiv{1202.3665}

\bibitem[{{Foreman-Mackey} {et~al.}(2013{\natexlab{b}}){Foreman-Mackey}, {Hogg}, {Lang}, \& {Goodman}}]{emcee}
---. 2013{\natexlab{b}}, \pasp, 125, 306, \dodoi{10.1086/670067}

\bibitem[{{Gaia Collaboration} \& {Brown et al.}(2018)}]{Gaia18}
{Gaia Collaboration}, \& {Brown et al.}, A.~G.~A. 2018, \aap, 616, A1.
\newblock \doarXiv{1804.09365}

\bibitem[{{Gaia Collaboration} {et~al.}(2016){Gaia Collaboration}, {Prusti}, {de Bruijne}, {Brown}, {Vallenari}, {Babusiaux}, {Bailer-Jones}, {Bastian}, {Biermann}, {Evans}, {Eyer}, {Jansen}, {Jordi}, {Klioner}, {Lammers}, {Lindegren}, {Luri}, {Mignard}, {Milligan}, {Panem}, {Poinsignon}, {Pourbaix}, {Randich}, {Sarri}, {Sartoretti}, {Siddiqui}, {Soubiran}, {Valette}, {van Leeuwen}, {Walton}, {Aerts}, {Arenou}, {Cropper}, {Drimmel}, {H{\o}g}, {Katz}, {Lattanzi}, {O'Mullane}, {Grebel}, {Holland}, {Huc}, {Passot}, {Bramante}, {Cacciari}, {Casta{\~n}eda}, {Chaoul}, {Cheek}, {De Angeli}, {Fabricius}, {Guerra}, {Hern{\'a}ndez}, {Jean-Antoine-Piccolo}, {Masana}, {Messineo}, {Mowlavi}, {Nienartowicz}, {Ord{\'o}{\~n}ez-Blanco}, {Panuzzo}, {Portell}, {Richards}, {Riello}, {Seabroke}, {Tanga}, {Th{\'e}venin}, {Torra}, {Els}, {Gracia-Abril}, {Comoretto}, {Garcia-Reinaldos}, {Lock}, {Mercier}, {Altmann}, {Andrae}, {Astraatmadja}, {Bellas-Velidis}, {Benson}, {Berthier}, {Blomme}, {Busso}, {Carry}, {Cellino}, {Clementini},
  {Cowell}, {Creevey}, {Cuypers}, {Davidson}, {De Ridder}, {de Torres}, {Delchambre}, {Dell'Oro}, {Ducourant}, {Fr{\'e}mat}, {Garc{\'\i}a-Torres}, {Gosset}, {Halbwachs}, {Hambly}, {Harrison}, {Hauser}, {Hestroffer}, {Hodgkin}, {Huckle}, {Hutton}, {Jasniewicz}, {Jordan}, {Kontizas}, {Korn}, {Lanzafame}, {Manteiga}, {Moitinho}, {Muinonen}, {Osinde}, {Pancino}, {Pauwels}, {Petit}, {Recio-Blanco}, {Robin}, {Sarro}, {Siopis}, {Smith}, {Smith}, {Sozzetti}, {Thuillot}, {van Reeven}, {Viala}, {Abbas}, {Abreu Aramburu}, {Accart}, {Aguado}, {Allan}, {Allasia}, {Altavilla}, {{\'A}lvarez}, {Alves}, {Anderson}, {Andrei}, {Anglada Varela}, {Antiche}, {Antoja}, {Ant{\'o}n}, {Arcay}, {Atzei}, {Ayache}, {Bach}, {Baker}, {Balaguer-N{\'u}{\~n}ez}, {Barache}, {Barata}, {Barbier}, {Barblan}, {Baroni}, {Barrado y Navascu{\'e}s}, {Barros}, {Barstow}, {Becciani}, {Bellazzini}, {Bellei}, {Bello Garc{\'\i}a}, {Belokurov}, {Bendjoya}, {Berihuete}, {Bianchi}, {Bienaym{\'e}}, {Billebaud}, {Blagorodnova}, {Blanco-Cuaresma}, {Boch},
  {Bombrun}, {Borrachero}, {Bouquillon}, {Bourda}, {Bouy}, {Bragaglia}, {Breddels}, {Brouillet}, {Br{\"u}semeister}, {Bucciarelli}, {Budnik}, {Burgess}, {Burgon}, {Burlacu}, {Busonero}, {Buzzi}, {Caffau}, {Cambras}, {Campbell}, {Cancelliere}, {Cantat-Gaudin}, {Carlucci}, {Carrasco}, {Castellani}, {Charlot}, {Charnas}, {Charvet}, {Chassat}, {Chiavassa}, {Clotet}, {Cocozza}, {Collins}, {Collins}, {Costigan}, {Crifo}, {Cross}, {Crosta}, {Crowley}, {Dafonte}, {Damerdji}, {Dapergolas}, {David}, {David}, {De Cat}, {de Felice}, {de Laverny}, {De Luise}, {De March}, {de Martino}, {de Souza}, {Debosscher}, {del Pozo}, {Delbo}, {Delgado}, {Delgado}, {di Marco}, {Di Matteo}, {Diakite}, {Distefano}, {Dolding}, {Dos Anjos}, {Drazinos}, {Dur{\'a}n}, {Dzigan}, {Ecale}, {Edvardsson}, {Enke}, {Erdmann}, {Escolar}, {Espina}, {Evans}, {Eynard Bontemps}, {Fabre}, {Fabrizio}, {Faigler}, {Falc{\~a}o}, {Farr{\`a}s Casas}, {Faye}, {Federici}, {Fedorets}, {Fern{\'a}ndez-Hern{\'a}ndez}, {Fernique}, {Fienga}, {Figueras}, {Filippi},
  {Findeisen}, {Fonti}, {Fouesneau}, {Fraile}, {Fraser}, {Fuchs}, {Furnell}, {Gai}, {Galleti}, {Galluccio}, {Garabato}, {Garc{\'\i}a-Sedano}, {Gar{\'e}}, {Garofalo}, {Garralda}, {Gavras}, {Gerssen}, {Geyer}, {Gilmore}, {Girona}, {Giuffrida}, {Gomes}, {Gonz{\'a}lez-Marcos}, {Gonz{\'a}lez-N{\'u}{\~n}ez}, {Gonz{\'a}lez-Vidal}, {Granvik}, {Guerrier}, {Guillout}, {Guiraud}, {G{\'u}rpide}, {Guti{\'e}rrez-S{\'a}nchez}, {Guy}, {Haigron}, {Hatzidimitriou}, {Haywood}, {Heiter}, {Helmi}, {Hobbs}, {Hofmann}, {Holl}, {Holland}, {Hunt}, {Hypki}, {Icardi}, {Irwin}, {Jevardat de Fombelle}, {Jofr{\'e}}, {Jonker}, {Jorissen}, {Julbe}, {Karampelas}, {Kochoska}, {Kohley}, {Kolenberg}, {Kontizas}, {Koposov}, {Kordopatis}, {Koubsky}, {Kowalczyk}, {Krone-Martins}, {Kudryashova}, {Kull}, {Bachchan}, {Lacoste-Seris}, {Lanza}, {Lavigne}, {Le Poncin-Lafitte}, {Lebreton}, {Lebzelter}, {Leccia}, {Leclerc}, {Lecoeur-Taibi}, {Lemaitre}, {Lenhardt}, {Leroux}, {Liao}, {Licata}, {Lindstr{\o}m}, {Lister}, {Livanou}, {Lobel}, {L{\"o}ffler},
  {L{\'o}pez}, {Lopez-Lozano}, {Lorenz}, {Loureiro}, {MacDonald}, {Magalh{\~a}es Fernandes}, {Managau}, {Mann}, {Mantelet}, {Marchal}, {Marchant}, {Marconi}, {Marie}, {Marinoni}, {Marrese}, {Marschalk{\'o}}, {Marshall}, {Mart{\'\i}n-Fleitas}, {Martino}, {Mary}, {Matijevi{\v{c}}}, {Mazeh}, {McMillan}, {Messina}, {Mestre}, {Michalik}, {Millar}, {Miranda}, {Molina}, {Molinaro}, {Molinaro}, {Moln{\'a}r}, {Moniez}, {Montegriffo}, {Monteiro}, {Mor}, {Mora}, {Morbidelli}, {Morel}, {Morgenthaler}, {Morley}, {Morris}, {Mulone}, {Muraveva}, {Musella}, {Narbonne}, {Nelemans}, {Nicastro}, {Noval}, {Ord{\'e}novic}, {Ordieres-Mer{\'e}}, {Osborne}, {Pagani}, {Pagano}, {Pailler}, {Palacin}, {Palaversa}, {Parsons}, {Paulsen}, {Pecoraro}, {Pedrosa}, {Pentik{\"a}inen}, {Pereira}, {Pichon}, {Piersimoni}, {Pineau}, {Plachy}, {Plum}, {Poujoulet}, {Pr{\v{s}}a}, {Pulone}, {Ragaini}, {Rago}, {Rambaux}, {Ramos-Lerate}, {Ranalli}, {Rauw}, {Read}, {Regibo}, {Renk}, {Reyl{\'e}}, {Ribeiro}, {Rimoldini}, {Ripepi}, {Riva}, {Rixon},
  {Roelens}, {Romero-G{\'o}mez}, {Rowell}, {Royer}, {Rudolph}, {Ruiz-Dern}, {Sadowski}, {Sagrist{\`a} Sell{\'e}s}, {Sahlmann}, {Salgado}, {Salguero}, {Sarasso}, {Savietto}, {Schnorhk}, {Schultheis}, {Sciacca}, {Segol}, {Segovia}, {Segransan}, {Serpell}, {Shih}, {Smareglia}, {Smart}, {Smith}, {Solano}, {Solitro}, {Sordo}, {Soria Nieto}, {Souchay}, {Spagna}, {Spoto}, {Stampa}, {Steele}, {Steidelm{\"u}ller}, {Stephenson}, {Stoev}, {Suess}, {S{\"u}veges}, {Surdej}, {Szabados}, {Szegedi-Elek}, {Tapiador}, {Taris}, {Tauran}, {Taylor}, {Teixeira}, {Terrett}, {Tingley}, {Trager}, {Turon}, {Ulla}, {Utrilla}, {Valentini}, {van Elteren}, {Van Hemelryck}, {van Leeuwen}, {Varadi}, {Vecchiato}, {Veljanoski}, {Via}, {Vicente}, {Vogt}, {Voss}, {Votruba}, {Voutsinas}, {Walmsley}, {Weiler}, {Weingrill}, {Werner}, {Wevers}, {Whitehead}, {Wyrzykowski}, {Yoldas}, {{\v{Z}}erjal}, {Zucker}, {Zurbach}, {Zwitter}, {Alecu}, {Allen}, {Allende Prieto}, {Amorim}, {Anglada-Escud{\'e}}, {Arsenijevic}, {Azaz}, {Balm}, {Beck}, {Bernstein},
  {Bigot}, {Bijaoui}, {Blasco}, {Bonfigli}, {Bono}, {Boudreault}, {Bressan}, {Brown}, {Brunet}, {Bunclark}, {Buonanno}, {Butkevich}, {Carret}, {Carrion}, {Chemin}, {Ch{\'e}reau}, {Corcione}, {Darmigny}, {de Boer}, {de Teodoro}, {de Zeeuw}, {Delle Luche}, {Domingues}, {Dubath}, {Fodor}, {Fr{\'e}zouls}, {Fries}, {Fustes}, {Fyfe}, {Gallardo}, {Gallegos}, {Gardiol}, {Gebran}, {Gomboc}, {G{\'o}mez}, {Grux}, {Gueguen}, {Heyrovsky}, {Hoar}, {Iannicola}, {Isasi Parache}, {Janotto}, {Joliet}, {Jonckheere}, {Keil}, {Kim}, {Klagyivik}, {Klar}, {Knude}, {Kochukhov}, {Kolka}, {Kos}, {Kutka}, {Lainey}, {LeBouquin}, {Liu}, {Loreggia}, {Makarov}, {Marseille}, {Martayan}, {Martinez-Rubi}, {Massart}, {Meynadier}, {Mignot}, {Munari}, {Nguyen}, {Nordlander}, {Ocvirk}, {O'Flaherty}, {Olias Sanz}, {Ortiz}, {Osorio}, {Oszkiewicz}, {Ouzounis}, {Palmer}, {Park}, {Pasquato}, {Peltzer}, {Peralta}, {P{\'e}turaud}, {Pieniluoma}, {Pigozzi}, {Poels}, {Prat}, {Prod'homme}, {Raison}, {Rebordao}, {Risquez}, {Rocca-Volmerange}, {Rosen},
  {Ruiz-Fuertes}, {Russo}, {Sembay}, {Serraller Vizcaino}, {Short}, {Siebert}, {Silva}, {Sinachopoulos}, {Slezak}, {Soffel}, {Sosnowska}, {Strai{\v{z}}ys}, {ter Linden}, {Terrell}, {Theil}, {Tiede}, {Troisi}, {Tsalmantza}, {Tur}, {Vaccari}, {Vachier}, {Valles}, {Van Hamme}, {Veltz}, {Virtanen}, {Wallut}, {Wichmann}, {Wilkinson}, {Ziaeepour}, \& {Zschocke}}]{2016A&A...595A...1G}
{Gaia Collaboration}, {Prusti}, T., {de Bruijne}, J.~H.~J., {et~al.} 2016, \aap, 595, A1, \dodoi{10.1051/0004-6361/201629272}

\bibitem[{{Gaia Collaboration} {et~al.}(2023){Gaia Collaboration}, {Vallenari}, {Brown}, {Prusti}, {de Bruijne}, {Arenou}, {Babusiaux}, {Biermann}, {Creevey}, {Ducourant}, {Evans}, {Eyer}, {Guerra}, {Hutton}, {Jordi}, {Klioner}, {Lammers}, {Lindegren}, {Luri}, {Mignard}, {Panem}, {Pourbaix}, {Randich}, {Sartoretti}, {Soubiran}, {Tanga}, {Walton}, {Bailer-Jones}, {Bastian}, {Drimmel}, {Jansen}, {Katz}, {Lattanzi}, {van Leeuwen}, {Bakker}, {Cacciari}, {Casta{\~n}eda}, {De Angeli}, {Fabricius}, {Fouesneau}, {Fr{\'e}mat}, {Galluccio}, {Guerrier}, {Heiter}, {Masana}, {Messineo}, {Mowlavi}, {Nicolas}, {Nienartowicz}, {Pailler}, {Panuzzo}, {Riclet}, {Roux}, {Seabroke}, {Sordo}, {Th{\'e}venin}, {Gracia-Abril}, {Portell}, {Teyssier}, {Altmann}, {Andrae}, {Audard}, {Bellas-Velidis}, {Benson}, {Berthier}, {Blomme}, {Burgess}, {Busonero}, {Busso}, {C{\'a}novas}, {Carry}, {Cellino}, {Cheek}, {Clementini}, {Damerdji}, {Davidson}, {de Teodoro}, {Nu{\~n}ez Campos}, {Delchambre}, {Dell'Oro}, {Esquej},
  {Fern{\'a}ndez-Hern{\'a}ndez}, {Fraile}, {Garabato}, {Garc{\'\i}a-Lario}, {Gosset}, {Haigron}, {Halbwachs}, {Hambly}, {Harrison}, {Hern{\'a}ndez}, {Hestroffer}, {Hodgkin}, {Holl}, {Jan{\ss}en}, {Jevardat de Fombelle}, {Jordan}, {Krone-Martins}, {Lanzafame}, {L{\"o}ffler}, {Marchal}, {Marrese}, {Moitinho}, {Muinonen}, {Osborne}, {Pancino}, {Pauwels}, {Recio-Blanco}, {Reyl{\'e}}, {Riello}, {Rimoldini}, {Roegiers}, {Rybizki}, {Sarro}, {Siopis}, {Smith}, {Sozzetti}, {Utrilla}, {van Leeuwen}, {Abbas}, {{\'A}brah{\'a}m}, {Abreu Aramburu}, {Aerts}, {Aguado}, {Ajaj}, {Aldea-Montero}, {Altavilla}, {{\'A}lvarez}, {Alves}, {Anders}, {Anderson}, {Anglada Varela}, {Antoja}, {Baines}, {Baker}, {Balaguer-N{\'u}{\~n}ez}, {Balbinot}, {Balog}, {Barache}, {Barbato}, {Barros}, {Barstow}, {Bartolom{\'e}}, {Bassilana}, {Bauchet}, {Becciani}, {Bellazzini}, {Berihuete}, {Bernet}, {Bertone}, {Bianchi}, {Binnenfeld}, {Blanco-Cuaresma}, {Blazere}, {Boch}, {Bombrun}, {Bossini}, {Bouquillon}, {Bragaglia}, {Bramante}, {Breedt},
  {Bressan}, {Brouillet}, {Brugaletta}, {Bucciarelli}, {Burlacu}, {Butkevich}, {Buzzi}, {Caffau}, {Cancelliere}, {Cantat-Gaudin}, {Carballo}, {Carlucci}, {Carnerero}, {Carrasco}, {Casamiquela}, {Castellani}, {Castro-Ginard}, {Chaoul}, {Charlot}, {Chemin}, {Chiaramida}, {Chiavassa}, {Chornay}, {Comoretto}, {Contursi}, {Cooper}, {Cornez}, {Cowell}, {Crifo}, {Cropper}, {Crosta}, {Crowley}, {Dafonte}, {Dapergolas}, {David}, {David}, {de Laverny}, {De Luise}, {De March}, {De Ridder}, {de Souza}, {de Torres}, {del Peloso}, {del Pozo}, {Delbo}, {Delgado}, {Delisle}, {Demouchy}, {Dharmawardena}, {Di Matteo}, {Diakite}, {Diener}, {Distefano}, {Dolding}, {Edvardsson}, {Enke}, {Fabre}, {Fabrizio}, {Faigler}, {Fedorets}, {Fernique}, {Fienga}, {Figueras}, {Fournier}, {Fouron}, {Fragkoudi}, {Gai}, {Garcia-Gutierrez}, {Garcia-Reinaldos}, {Garc{\'\i}a-Torres}, {Garofalo}, {Gavel}, {Gavras}, {Gerlach}, {Geyer}, {Giacobbe}, {Gilmore}, {Girona}, {Giuffrida}, {Gomel}, {Gomez}, {Gonz{\'a}lez-N{\'u}{\~n}ez},
  {Gonz{\'a}lez-Santamar{\'\i}a}, {Gonz{\'a}lez-Vidal}, {Granvik}, {Guillout}, {Guiraud}, {Guti{\'e}rrez-S{\'a}nchez}, {Guy}, {Hatzidimitriou}, {Hauser}, {Haywood}, {Helmer}, {Helmi}, {Sarmiento}, {Hidalgo}, {Hilger}, {H{\l}adczuk}, {Hobbs}, {Holland}, {Huckle}, {Jardine}, {Jasniewicz}, {Jean-Antoine Piccolo}, {Jim{\'e}nez-Arranz}, {Jorissen}, {Juaristi Campillo}, {Julbe}, {Karbevska}, {Kervella}, {Khanna}, {Kontizas}, {Kordopatis}, {Korn}, {K{\'o}sp{\'a}l}, {Kostrzewa-Rutkowska}, {Kruszy{\'n}ska}, {Kun}, {Laizeau}, {Lambert}, {Lanza}, {Lasne}, {Le Campion}, {Lebreton}, {Lebzelter}, {Leccia}, {Leclerc}, {Lecoeur-Taibi}, {Liao}, {Licata}, {Lindstr{\o}m}, {Lister}, {Livanou}, {Lobel}, {Lorca}, {Loup}, {Madrero Pardo}, {Magdaleno Romeo}, {Managau}, {Mann}, {Manteiga}, {Marchant}, {Marconi}, {Marcos}, {Marcos Santos}, {Mar{\'\i}n Pina}, {Marinoni}, {Marocco}, {Marshall}, {Martin Polo}, {Mart{\'\i}n-Fleitas}, {Marton}, {Mary}, {Masip}, {Massari}, {Mastrobuono-Battisti}, {Mazeh}, {McMillan}, {Messina}, {Michalik},
  {Millar}, {Mints}, {Molina}, {Molinaro}, {Moln{\'a}r}, {Monari}, {Mongui{\'o}}, {Montegriffo}, {Montero}, {Mor}, {Mora}, {Morbidelli}, {Morel}, {Morris}, {Muraveva}, {Murphy}, {Musella}, {Nagy}, {Noval}, {Oca{\~n}a}, {Ogden}, {Ordenovic}, {Osinde}, {Pagani}, {Pagano}, {Palaversa}, {Palicio}, {Pallas-Quintela}, {Panahi}, {Payne-Wardenaar}, {Pe{\~n}alosa Esteller}, {Penttil{\"a}}, {Pichon}, {Piersimoni}, {Pineau}, {Plachy}, {Plum}, {Poggio}, {Pr{\v{s}}a}, {Pulone}, {Racero}, {Ragaini}, {Rainer}, {Raiteri}, {Rambaux}, {Ramos}, {Ramos-Lerate}, {Re Fiorentin}, {Regibo}, {Richards}, {Rios Diaz}, {Ripepi}, {Riva}, {Rix}, {Rixon}, {Robichon}, {Robin}, {Robin}, {Roelens}, {Rogues}, {Rohrbasser}, {Romero-G{\'o}mez}, {Rowell}, {Royer}, {Ruz Mieres}, {Rybicki}, {Sadowski}, {S{\'a}ez N{\'u}{\~n}ez}, {Sagrist{\`a} Sell{\'e}s}, {Sahlmann}, {Salguero}, {Samaras}, {Sanchez Gimenez}, {Sanna}, {Santove{\~n}a}, {Sarasso}, {Schultheis}, {Sciacca}, {Segol}, {Segovia}, {S{\'e}gransan}, {Semeux}, {Shahaf}, {Siddiqui}, {Siebert},
  {Siltala}, {Silvelo}, {Slezak}, {Slezak}, {Smart}, {Snaith}, {Solano}, {Solitro}, {Souami}, {Souchay}, {Spagna}, {Spina}, {Spoto}, {Steele}, {Steidelm{\"u}ller}, {Stephenson}, {S{\"u}veges}, {Surdej}, {Szabados}, {Szegedi-Elek}, {Taris}, {Taylor}, {Teixeira}, {Tolomei}, {Tonello}, {Torra}, {Torra}, {Torralba Elipe}, {Trabucchi}, {Tsounis}, {Turon}, {Ulla}, {Unger}, {Vaillant}, {van Dillen}, {van Reeven}, {Vanel}, {Vecchiato}, {Viala}, {Vicente}, {Voutsinas}, {Weiler}, {Wevers}, {Wyrzykowski}, {Yoldas}, {Yvard}, {Zhao}, {Zorec}, {Zucker}, \& {Zwitter}}]{2023A&A...674A...1G}
{Gaia Collaboration}, {Vallenari}, A., {Brown}, A.~G.~A., {et~al.} 2023, \aap, 674, A1, \dodoi{10.1051/0004-6361/202243940}

\bibitem[{Harris {et~al.}(2020)Harris, Millman, van~der Walt, Gommers, Virtanen, Cournapeau, Wieser, Taylor, Berg, Smith, Kern, Picus, Hoyer, van Kerkwijk, Brett, Haldane, del R{\'{i}}o, Wiebe, Peterson, G{\'{e}}rard-Marchant, Sheppard, Reddy, Weckesser, Abbasi, Gohlke, \& Oliphant}]{harris2020array}
Harris, C.~R., Millman, K.~J., van~der Walt, S.~J., {et~al.} 2020, Nature, 585, 357, \dodoi{10.1038/s41586-020-2649-2}

\bibitem[{{Hills}(1988)}]{hills88}
{Hills}, J.~G. 1988, \nat, 331, 687, \dodoi{10.1038/331687a0}

\bibitem[{Hunter(2007)}]{Hunter07}
Hunter, J.~D. 2007, Computing in Science \& Engineering, 9, 90, \dodoi{10.1109/MCSE.2007.55}

\bibitem[{{Koposov} {et~al.}(2020){Koposov}, {Boubert}, {Li}, {Erkal}, {Da Costa}, {Zucker}, {Ji}, {Kuehn}, {Lewis}, {Mackey}, {Simpson}, {Shipp}, {Wan}, {Belokurov}, {Bland-Hawthorn}, {Martell}, {Nordlander}, {Pace}, {De Silva}, {Wang}, \& {S5 Collaboration}}]{koposov20}
{Koposov}, S.~E., {Boubert}, D., {Li}, T.~S., {et~al.} 2020, \mnras, 491, 2465.
\newblock \doarXiv{1907.11725}

\bibitem[{{Kuijken} \& {Rich}(2002)}]{kuijken02}
{Kuijken}, K., \& {Rich}, R.~M. 2002, \aj, 124, 2054.
\newblock \doarXiv{astro-ph/0207116}

\bibitem[{{Li} {et~al.}(2021){Li}, {Luo}, {Lu}, {Zhang}, {Li}, {Wang}, {Zuo}, {Xiang}, {Ting}, {Marchetti}, {Li}, {Wang}, {Zhang}, {Hattori}, {Zhao}, {Zhang}, \& {Zhao}}]{li21}
{Li}, Y.-B., {Luo}, A.~L., {Lu}, Y.-J., {et~al.} 2021, \apjs, 252, 3, \dodoi{10.3847/1538-4365/abc16e}

\bibitem[{{Luna} {et~al.}(2019){Luna}, {Minniti}, \& {Alonso-Garc{\'\i}a}}]{luna19}
{Luna}, A., {Minniti}, D., \& {Alonso-Garc{\'\i}a}, J. 2019, \apjl, 887, L39.
\newblock \doarXiv{1912.02129}

\bibitem[{{Luri} {et~al.}(2018){Luri}, {Brown}, {Sarro}, {Arenou}, {Bailer-Jones}, {Castro-Ginard}, {de Bruijne}, {Prusti}, {Babusiaux}, \& {Delgado}}]{Luri2018}
{Luri}, X., {Brown}, A.~G.~A., {Sarro}, L.~M., {et~al.} 2018, \aap, 616, A9, \dodoi{10.1051/0004-6361/201832964}

\bibitem[{{Marchetti}(2021)}]{marchetti21}
{Marchetti}, T. 2021, \mnras, 503, 1374.
\newblock \doarXiv{2012.02123}

\bibitem[{{Marchetti} {et~al.}(2019){Marchetti}, {Rossi}, \& {Brown}}]{marchetti19}
{Marchetti}, T., {Rossi}, E.~M., \& {Brown}, A.~G.~A. 2019, \mnras, 490, 157, \dodoi{10.1093/mnras/sty2592}

\bibitem[{{Minniti} {et~al.}(2010){Minniti}, {Lucas}, {Emerson}, {Saito}, {Hempel}, {Pietrukowicz}, {Ahumada}, {Alonso}, {Alonso-Garcia}, {Arias}, {Bandyopadhyay}, {Barb{\'a}}, {Barbuy}, {Bedin}, {Bica}, {Borissova}, {Bronfman}, {Carraro}, {Catelan}, {Clari{\'a}}, {Cross}, {de Grijs}, {D{\'e}k{\'a}ny}, {Drew}, {Fari{\~n}a}, {Feinstein}, {Fern{\'a}ndez Laj{\'u}s}, {Gamen}, {Geisler}, {Gieren}, {Goldman}, {Gonzalez}, {Gunthardt}, {Gurovich}, {Hambly}, {Irwin}, {Ivanov}, {Jord{\'a}n}, {Kerins}, {Kinemuchi}, {Kurtev}, {L{\'o}pez-Corredoira}, {Maccarone}, {Masetti}, {Merlo}, {Messineo}, {Mirabel}, {Monaco}, {Morelli}, {Padilla}, {Palma}, {Parisi}, {Pignata}, {Rejkuba}, {Roman-Lopes}, {Sale}, {Schreiber}, {Schr{\"o}der}, {Smith}, {}, {Soto}, {Tamura}, {Tappert}, {Thompson}, {Toledo}, {Zoccali}, \& {Pietrzynski}}]{2010NewA...15..433M}
{Minniti}, D., {Lucas}, P.~W., {Emerson}, J.~P., {et~al.} 2010, \na, 15, 433, \dodoi{10.1016/j.newast.2009.12.002}

\bibitem[{{Monari} {et~al.}(2018){Monari}, {Famaey}, {Carrillo}, {Piffl}, {Steinmetz}, {Wyse}, {Anders}, {Chiappini}, \& {Jan{\ss}en}}]{monari18}
{Monari}, G., {Famaey}, B., {Carrillo}, I., {et~al.} 2018, \aap, 616, L9, \dodoi{10.1051/0004-6361/201833748}

\bibitem[{{Schechter} {et~al.}(1993){Schechter}, {Mateo}, \& {Saha}}]{schechter93}
{Schechter}, P.~L., {Mateo}, M., \& {Saha}, A. 1993, \pasp, 105, 1342, \dodoi{10.1086/133316}

\bibitem[{{Sumi} {et~al.}(2004){Sumi}, {Wu}, {Udalski}, {Szyma{\'n}ski}, {Kubiak}, {Pietrzy{\'n}ski}, {Soszy{\'n}ski}, {Wo{\'z}niak}, {{\.Z}ebru{\'n}}, {Szewczyk}, \& {Wyrzykowski}}]{2004MNRAS.348.1439S}
{Sumi}, T., {Wu}, X., {Udalski}, A., {et~al.} 2004, \mnras, 348, 1439, \dodoi{10.1111/j.1365-2966.2004.07457.x}

\bibitem[{{Szymanski} \& {Udalski}(1993)}]{Szymanski93}
{Szymanski}, M., \& {Udalski}, A. 1993, \actaa, 43, 91

\bibitem[{{Tonry} {et~al.}(2012){Tonry}, {Stubbs}, {Kilic}, {Flewelling}, {Deacon}, {Chornock}, {Berger}, {Burgett}, {Chambers}, {Kaiser}, {Kudritzki}, {Hodapp}, {Magnier}, {Morgan}, {Price}, \& {Wainscoat}}]{2012ApJ...745...42T}
{Tonry}, J.~L., {Stubbs}, C.~W., {Kilic}, M., {et~al.} 2012, \apj, 745, 42, \dodoi{10.1088/0004-637X/745/1/42}

\bibitem[{{Udalski} {et~al.}(1997){Udalski}, {Kubiak}, \& {Szymanski}}]{udalski97}
{Udalski}, A., {Kubiak}, M., \& {Szymanski}, M. 1997, \actaa, 47, 319.
\newblock \doarXiv{astro-ph/9710091}

\bibitem[{{Udalski} {et~al.}(1993){Udalski}, {Szymanski}, {Kaluzny}, {Kubiak}, {Krzeminski}, {Mateo}, {Preston}, \& {Paczynski}}]{udalski93}
{Udalski}, A., {Szymanski}, M., {Kaluzny}, J., {et~al.} 1993, \actaa, 43, 289

\bibitem[{{Udalski} {et~al.}(2015){Udalski}, {Szyma{\'n}ski}, \& {Szyma{\'n}ski}}]{udalski15}
{Udalski}, A., {Szyma{\'n}ski}, M.~K., \& {Szyma{\'n}ski}, G. 2015, \actaa, 65, 1.
\newblock \doarXiv{1504.05966}

\bibitem[{{Williams} {et~al.}(2017){Williams}, {Belokurov}, {Casey}, \& {Evans}}]{williams17}
{Williams}, A.~A., {Belokurov}, V., {Casey}, A.~R., \& {Evans}, N.~W. 2017, \mnras, 468, 2359, \dodoi{10.1093/mnras/stx508}

\bibitem[{{Yu} \& {Tremaine}(2003)}]{yu03}
{Yu}, Q., \& {Tremaine}, S. 2003, \apj, 599, 1129.
\newblock \doarXiv{astro-ph/0309084}

\bibitem[{{Ziegerer} {et~al.}(2015){Ziegerer}, {Volkert}, {Heber}, {Irrgang}, {G{\"a}nsicke}, \& {Geier}}]{Ziegerer15}
{Ziegerer}, E., {Volkert}, M., {Heber}, U., {et~al.} 2015, \aap, 576, L14, \dodoi{10.1051/0004-6361/201526052}

\end{thebibliography}
\bibliographystyle{aasjournal}

\end{document}